\documentclass[manuscript]{aastex631}

\newcommand\dt{$\Delta t_{\rm O-X}$ }
\def\approxgt{\ifmmode \rlap{$>$}{}_{{}_{{}_{\textstyle\sim}}} \else%
$\rlap{$>$}{}_{{}_{{}_{\textstyle\sim}}}$\fi} 
\def\approxlt{\ifmmode \rlap{$<$}{}_{{}_{{}_{\textstyle\sim}}} \else%
$\rlap{$<$}{}_{{}_{{}_{\textstyle\sim}}}$\fi}
\usepackage{color}

\usepackage[T1]{fontenc}

\shorttitle{Delayed X-ray flares in optically-selected TDEs}
\shortauthors{Hayasaki \& Jonker}

\begin{document}




\title{On the origin of late-time X-ray flares in UV/optically-selected tidal disruption events}

\correspondingauthor{Kimitake Hayasaki}
\email{kimi@cbnu.ac.kr}

\author{Kimitake Hayasaki}
\affiliation{Department of Astronomy and Space Science, Chungbuk National University, Cheongju 361-763, Korea}
\author{Peter G.~Jonker}
\affiliation{Department of Astrophysics/IMAPP, Radboud University, P.O. Box 9010, 6500 GL, Nijmegen, The Netherlands}
\affiliation{SRON, Netherlands Institute for Space Research, Sorbonnelaan 2, 3584 CA, Utrecht, The Netherlands}






\begin{abstract}
We propose a model to explain the time delay between the peak of the optical and X-ray luminosity, \dt hereafter, in UV/optically-selected tidal disruption events (TDEs). The following picture explains the observed \dt in several TDEs as a consequence of the circularization and disk accretion processes as long as the sub-Eddington accretion. At the beginning of the circularization, the fallback debris is thermalized by the self-crossing shock caused by relativistic precession, providing the peak optical emission. During the circularization process, the mass fallback rate decreases with time to form a ring around the supermassive black hole (SMBH). The formation timescale corresponds to the circularization timescale of the most tightly bound debris, which is less than a year to several decades, depending mostly on the penetration factor, the circularization efficiency, and the black hole mass. The ring will subsequently evolve viscously over the viscous diffusion time. We find that it accretes onto the SMBH on a fraction of the viscous timescale, which is $2$ years for given typical parameters, leading to X-ray emission at late times. The resultant \dt\,is given by the sum of the circularization timescale and the accretion timescale and significantly decreases with increasing penetration factor to several to $\sim10$ years typically. Since the X-ray luminosity substantially decreases as the viewing angle between the normal to the disk plane and line-of-sight increases from $0^\circ$ to $90^\circ$, a low late-time X-ray luminosity can be explained by an edge-on view. We also discuss the super-Eddington accretion scenario, where \dt\,is dominated by the circularization timescale.
%
%
\end{abstract}

%

%
\section{Introduction}
\label{sec:intro}
%

%


%
%
Tidal disruption events (TDEs) are becoming a key phenomenon in searching 
for dormant supermassive black holes (SMBHs) at the centers of the inactive galaxies. 
TDEs occur when a star approaches close enough to the SMBH to be ripped apart 
by its tidal force. The subsequent accretion of a fraction of the mass from the tidally disrupted star causes a characteristic 
flare with a high luminosity for 
a time scale of weeks to months to years, in exceptional cases \citep{1988Natur.333..523R,1989ApJ...346L..13E,1989IAUS..136..543P,2009MNRAS.392..332L}. For SMBH masses $\approxlt 10^7$~M$_\odot$ the luminosity can even exceed the Eddington luminosity (\citealt{2013ApJ...767...25G}).

Recent multi-wavelength observations have revealed a diverse set of properties for TDEs. They can roughly be divided into two categories: thermal TDEs without a strong relativistic jet and non-thermal TDEs with a relativistic jet that probably has a relatively low inclination angle with respect to our line of sight, i.e., Jetted TDEs. The inferred event rate of thermal TDEs is $10^{-5}-10^{-4}$ per year per galaxy (see \citealt{2020SSRv..216...35S} for a recent review), whereas that of the jetted TDEs is much lower as evidenced by the detection of only three jetted TDEs despite their much larger (beamed) luminosity  \citep{2011Sci...333..199L,2011Natur.476..421B,2012ApJ...753...77C,2015MNRAS.452.4297B}. While some of the thermal TDEs shine brightly only in soft-X-ray wavebands (i.e., soft-X-ray TDEs) \citep[see][for a recent review]{2021SSRv..217...18S}, others are bright mainly in optical/UV wavebands (i.e., optical/UV TDEs; see \citealt[][]{2020SSRv..216..124V} for a recent review). In addition, there is a growing number of optical, radio, {\it and} X-ray bright TDEs (e.g., \citealt{2016MNRAS.455.2918H,2016Sci...351...62V,2017ApJ...851L..47G,2019MNRAS.488.4816W,2020MNRAS.499..482N,2020arXiv200505340S,2021MNRAS.tmp..840C}).
Interestingly, one of these, AT2019dsg, might also be responsible for (possible) neutrino emission \citep{2020arXiv200505340S,2021NatAs.tmp...40H}. It is proposed that the observed diversity of thermal TDEs can be explained, in part, by the viewing angle of the observer relative to the orientation of the disk angular momentum \citep{2018ApJ...859L..20D}. 

In another thermal TDE, ASASSN-15oi, the X-ray emission rose 1 year after the optical/UV peak as observed by {\it Swift} and {\it XMM-Newon} \citep{2017ApJ...851L..47G}. It has been proposed that this late rise in X-ray emission in ASASSN-15oi was related to the accretion disk becoming less puffed-up with time due to a decrease in mass accretion rate and/or the settling of the accretion disk in the BH equatorial plane (\citealt{2020ApJ...897...80W}). 
Besides, ASASSN-15lh is a highly luminous optical/UV transient regarded as a TDE candidate to explain the observational properties. The observed UV rebrightening at $\sim120\,{\rm days}$ after the first UV peak is interpreted by viscously-delayed, reprocessed, accretion onto the SMBH \citep{2016NatAs...1E...2L}. 
Recently, using {\it Chandra} observations \citet[][J20 hereafter]{2020ApJ...889..166J} have detected X-ray emission at late times ($\approx 4.5-9$~yr) from a few optically-selected TDEs. The period (i.e., \dt) between the optical and X-ray detections is several years on average, although the X-ray coverage of the optical light curve is sparse so a shorter delay time is possible as well in many cases.  In J20 it was mentioned that the \dt will depend on the TDE parameters such as $\beta$, and the mass and spin of the black hole. 
Instead of the geometrical effect invoked by \citet{2020ApJ...897...80W} to explain ASASSN-15oi's behavior, which could indeed play a role as well in individual systems, in this paper, we aim to explain and quantify the \dt\,in TDEs in terms of a difference in the emission mechanisms brought about by the circularization and disk accretion processes.

%
\section{Viscous diffusion of the circularized disk}
\label{sec:2}
%
The tidal disruption radius expressed in units of the Schwarzschild radius is given by
\begin{equation}
\frac{r_{\rm t}}{r_{\rm S}}=\left(\frac{M_{\rm bh}}{m_{*}}\right)^{1/3}\frac{r_{*}}{r_{\rm S}}
\approx
24.5\,
M_{\rm bh,6}^{-2/3}\,m_{*,1}^{-1/3}\,r_{*,1}
,
\end{equation}
where $M_{\rm bh}$ is mass of the central SMBH, $m_*$ and $r_*$ 
are the stellar mass and radius, $r_{\rm S}=2GM_{\rm bh}/c^2$ is the 
Schwarzschild radius of the SMBH, $c$ is the speed of light, and we adopt $M_{\rm bh,6}=M_{\rm bh}/10^6M_\odot$, $m_{*,1}=m_*/M_\odot$, and $r_{*,1}=r_*/R_\odot$ as fiducial values throughout the paper \footnote{As $r_{\rm t}$ is expressed in units of the Schwarzschild radius, which is in itself a function of SMBH mass, the dependence of $r_{\rm t}/r_{\rm S}$ on SMBH mass goes as M$_{\rm bh,6}$ to the power of $-2/3$.}, unless otherwise noted. We furthermore take $\beta=1$ as fiducial value, where $\beta$ is the penetration factor, which is defined as the ratio between the tidal disruption and the orbital pericenter radii of the disrupted star. After the tidal disruption of a star, the stellar debris falls back towards the SMBH and the stream will self-interact. In a so called self-interaction shock orbital energy of the stream is converted into thermal energy. 

It is still debated if, and if so, how all the stellar debris efficiently circularizes by the stream-stream collision. Some hydrodynamical simulations show that the TDE disk retains a significantly elliptical shape because the orbital energy is not dissipated efficiently enough to reduce the eccentricity of the entire disk to zero in a reasonable time \citep{2014ApJ...783...23G,2015ApJ...804...85S,2016MNRAS.458.4250S}. \cite{2020MNRAS.492..686L} show that a significant fraction of the debris can become unbound causing an outflow from the self-interaction region. Nevertheless, the debris that remains bound eventually contributes to the accretion flow around the SMBH.
This part of the debris stream will finally be circularized by energy dissipation, leading to the formation of a small, initially ring-like, accretion disk around the black hole \citep{2013MNRAS.434..909H,2016MNRAS.455.2253B,2016MNRAS.461.3760H}. Note that, in an inefficient debris circularization case, the subsequent fallback material interacts with the outer elliptical debris so that their effect on the subsequent evolution of the initial ring is negligible. Angular momentum conservation allows us to estimate the circularization radius of the stellar debris, which is given by 
\begin{eqnarray}
r_{\rm{c}}
=(1+e_*)r_{\rm p},
=\frac{1+e_*}{\beta}r_{\rm{t}},
\label{eq:rc}
\end{eqnarray}
where $e_*$ is the orbital eccentricity of the stellar orbit, $r_{\rm p}=r_{\rm t}/\beta$ is the pericenter distance radius. If debris circularization takes place only through dissipation at the self-interaction shock, 
the circularization timescale for the non-magnetized, most tightly bound debris can be estimated 
based on the ballistic approximation \citep{2017MNRAS.464.2816B} as
\begin{eqnarray}
t_{\rm circ}
&&
\approx8.3\,\eta^{-1}\,\beta^{-3}M_{\rm bh,6}^{-5/3}\,t_{\rm mtb}
\sim 0.93\,
\left(\frac{\eta}{1.0}\right)^{-1}
\beta^{-3}
M_{\rm bh,6}^{-7/6}\,m_{*,1}^{-1}\,r_{*,1}^{3/2}\,\,{\rm yr}\,,
\label{eq:tcirc}
\end{eqnarray}
where the orbital period of the stellar debris on the most tightly bound orbit:
\begin{equation}
t_{\rm mtb}=\frac{\pi}{\sqrt{2}}\frac{1}{\Omega_{*}}\left(\frac{M_{\rm bh}}{m_*}\right)^{1/2}
\approx
0.11\,
M_{\rm bh,6}^{1/2}\,m_{*,1}^{-1}\,r_{*,1}^{3/2}\,\,{\rm yr}
\label{eq:mtb}
\end{equation}
and $\Omega_{*}=\sqrt{Gm_{*}/r_{*}^{3}}$ is the dynamical angular frequency of the star, and we introduce $\eta\,(\le1)$ as the circularization efficiency which represents how efficiently the kinetic energy at the stream-stream collision is dissipated and the most efficient ($\eta=1$) case corresponds to that of \cite{2017MNRAS.464.2816B}. Note that $t_{\rm circ}$ is not the circularization timescale of all the stellar debris. Our interest here is in the circularization timescale and radius of the most tightly bound debris because the accretion of this debris contributes most to the delayed X-ray peak luminosity in terms of the emitted energy.

The mass fallback rate is given by
\begin{eqnarray}
\dot{M}=
\zeta\frac{m_{*}}{t_{\rm mtb}}\left(\frac{t}{t_{\rm mtb}}\right)^{-5/3}
\sim1.9\times10^{25}
\,
M_{\rm bh,6}^{-1/2}\,m_{*,1}^{2}\,r_{*,1}^{-3/2}
\left(\frac{\zeta}{1/3}\right)
\left(\frac{t}{t_{\rm mtb}}\right)^{-5/3}
\,\,{\rm g\,s^{-1}},
\label{eq:mdot}
\end{eqnarray}
where $\zeta=1/3$ is for the standard case \citep{1989ApJ...346L..13E,1989IAUS..136..543P}. 
If a star on a slightly hyperbolic orbit is tidally disrupted by a SMBH, the resultant stellar debris is more loosely bound than the standard parabolic case, reducing the mass fallback rate significantly \citep{2018ApJ...855..129H}. These hyperbolic TDEs have $\zeta < 1/3$, and the power-law index of the temporal evolution is slightly larger than $-5/3$ \citep{2018ApJ...855..129H,2020ApJ...900....3P}. 
In the discussion here, the detailed slope of the mass fallback rate is not important, therefore, we do not parametrize the temporal power-law index. 

The process responsible for the high optical luminosity in TDEs is still a matter of debate. One of the promising models is 
the shock-powered scenario where most of the observed optical emission is released around the self-interaction shock \citep{2015ApJ...804...85S,2015ApJ...806..164P,2020ApJ...904...73R}. By using equations (\ref{eq:mtb}) and (\ref{eq:mdot}), the peak optical luminosity is estimated to be at most
\begin{eqnarray}
L_{\rm shock}
&&
\sim\frac{GM_{\rm bh}\dot{M}(t_{\rm mtb})
}{a_{\rm mtb}}
\sim
7.1\times10^{43}\,
\left(\frac{\zeta}{1/3}\right)
M_{\rm bh,6}^{-1/6}\,m_{*,1}^{8/3}\,r_{*,1}^{-5/2}
\,\,{\rm erg/s},
\label{eq:lshock}
\end{eqnarray}
where $a_{\rm mtb}$ is the orbital semi-major axis of the most tightly bound debris. We can evaluate $a_{\rm mtb}$ by equating the debris orbital energy $|-GM_{\rm bh}/2a_{\rm mtb}|$ with the tidal potential $GM_{\rm bh}r_*/r_{\rm t}^2$ as
\begin{equation}
a_{\rm mtb}=\frac{1}{2}\frac{r_{\rm t}^2}{r_{*}}\sim3.5\times10^{14}\,M_{\rm bh,6}^{2/3}\,m_{*,1}^{-2/3}\,r_{*,1}\,\,{\rm cm},
\end{equation}
where we neglect the effect that the internal structure of the star will have on the spread of the orbital energies over the debris. Note that $L_{\rm shock}$ can be more than one order of magnitude lower if the heat produced by the shock is liberated radiatively inefficiently \citep{2016ApJ...830..125J}.
Since the energy liberated in the shock according to equation (\ref{eq:lshock}) is too low for the entire debris to be fully circularized, the remaining debris could form an eccentric disk \citep{2017MNRAS.467.1426S,2017MNRAS.472L..99L,2020MNRAS.499.5562Z,2021MNRAS.500.4110L}. However, even for an eccentric disk, the bolometric luminosity is comparable to $L_{\rm shock}$ \citep{2020MNRAS.499.5562Z}.

Substituting equation (\ref{eq:tcirc}) into equation (\ref{eq:mdot}), we obtain the mass fallback rate 
after the most tightly bound debris is circularized:
\begin{eqnarray}
\frac{
\dot{M}_{\rm c}
}{\dot{M}_{\rm Edd}}
\sim0.85
\left(\frac{\eta}{0.1}\right)^{5/3}
\left(\frac{\zeta}{1/3}\right)
\beta^{5}
M_{\rm bh,6}^{23/18}\,m_{*,1}^{2}\,r_{*,1}^{-3/2}
\,\,{\rm g\,s^{-1}},
\label{eq:mdotc}
\end{eqnarray}
where $\dot{M}_{\rm Edd}=L_{\rm Edd}/c^2\simeq1.4\times10^{23}\,M_{\rm bh,6}\,{\rm g/s}$ is the Eddington accretion rate and 
\begin{equation}
L_{\rm Edd}=\frac{4\pi{GM_{\rm bh}}m_{p}c}{\sigma_{\rm T}}\simeq1.3\times10^{44}\,M_{\rm bh,6}
\,\,{\rm erg/s}\,
\end{equation}
is the Eddington luminosity with the proton mass $m_{\rm p}$ and the Thomson scattering cross section $\sigma_{\rm T}$. 
For the given black hole and stellar masses and the stellar radius, we obtain the condition that $\dot{M}_{\rm c}$ is smaller than $\dot{M}_{\rm Edd}$ to be sub-Eddington rate as
\begin{eqnarray}
\zeta\,\eta^{5/3}\,\beta^5\,m_{*,1}^{2}\,r_{*,1}^{-3/2}\,M_{\rm bh,6}^{23/18}\,
\lesssim8.5\times10^{-3}
\label{eq:prange}
\end{eqnarray}
Our model noted below is applicable for this parameter range.

The viscous timescale is estimated through the $\alpha$ viscosity prescription (\citealt{1973A&A....24..337S}) as
\begin{equation}
t_{\rm vis}(r)=\frac{r_{\rm }^2}{\nu_{\rm vis }}\approx\frac{1}{\alpha}\frac{1}{\Omega}
\left(\frac{H}{r}\right)^{-2}
\sim
14
\left(\frac{\alpha}{0.1}\right)^{-1}
\left(\frac{H/r}{0.01}\right)^{-2}
\left(\frac{r}{r_{\rm c}}\right)^{3/2}
\left(\frac{1+e_*}{2}\right)^{3/2}
\beta^{-3/2}
m_{*,1}^{-1/2}
r_{*,1}^{3/2}
\,\,\rm yr\,,
\label{eq:tvis}
\end{equation}
where $\nu_{\rm vis}=\alpha{c_{\rm s}}H$, $\alpha$ is the Shakura-Sunyaev viscosity parameter, $c_{\rm s}$ is the sound speed in the disk, $H$ is the disk scale height, and $\Omega=\sqrt{GM_{\rm bh}/r^3}$ is a Keplerian angular frequency of the disk. Adopting the standard disk model \citep{1973A&A....24..337S}, the square of the disk scale-height 
at $r_{\rm c}$ is estimated to be
\begin{eqnarray}
\left(\frac{H}{r_{\rm c}}\right)^2
\sim
1.0\times10^{-4}\,
\left(\frac{\alpha}{0.1}\right)^{-1/5}
\left(\frac{\zeta}{1/3}\right)^{2/5}
\left(\frac{1+e_*}{2}\right)^{1/10}
\beta^{19/10}
M_{\rm bh,6}^{11/45}
m_{*,1}^{23/30}
r_{*,1}^{-1/2}.
\label{eq:hr2}
\end{eqnarray}
Substituting equation (\ref{eq:hr2}) into (\ref{eq:tvis}), we obtain the viscous diffusion timescale at $r_{\rm c}$ as
\begin{equation}
t_{\rm c}
=
\frac{4}{3}t_{\rm vis}(r_{\rm c})
\approx
19
\left(\frac{\alpha}{0.1}\right)^{-4/5}
\left(\frac{\zeta}{1/3}\right)^{-2/5}
\left(\frac{1+e_*}{2}\right)^{7/5}
\beta^{-17/5}
M_{\rm bh,6}^{-11/45}
m_{*,1}^{-19/15}
r_{*,1}^{2}
\,\,\rm yr\,
.
\label{eq:tc}
\end{equation} 
Because $t_{\rm c}$ is much longer than the circularization timescale of the most tightly bound debris, 
the subsequent disk evolves viscously.

%
\subsection{Evolution and structure of a time-dependent accretion disk}
%

Let us describe the subsequent disk evolution. We assume that the disk is one-dimensional (1D), axisymmetric.
Mass and angular momentum conservation laws of the viscous accretion disk for a point-mass gravitational potential
provide the equation for the surface density evolution as (e.g., see \citealt{2002apa..book.....F,2008bhad.book.....K} and references therein)
\begin{eqnarray}
\frac{\partial\Sigma}{\partial t}=-\frac{2}{r}\frac{\partial}{\partial r}
\left[
\left(\frac{r}{GM}\right)^{1/2}
\frac{\partial}{\partial r}
(r^2\mathcal{T}_{r\phi})
\right],
\label{eq:diskevo}
\end{eqnarray}
where $\mathcal{T}_{r\phi}=\nu_{\rm vis}\Sigma{r}{d\Omega/dr}=-(3/2)\nu_{\rm vis}\Sigma\Omega$ 
is the viscous stress tensor of the disk in Keplerian rotation. 

\begin{figure}[!htbp]
\centering
\includegraphics[width=14cm]{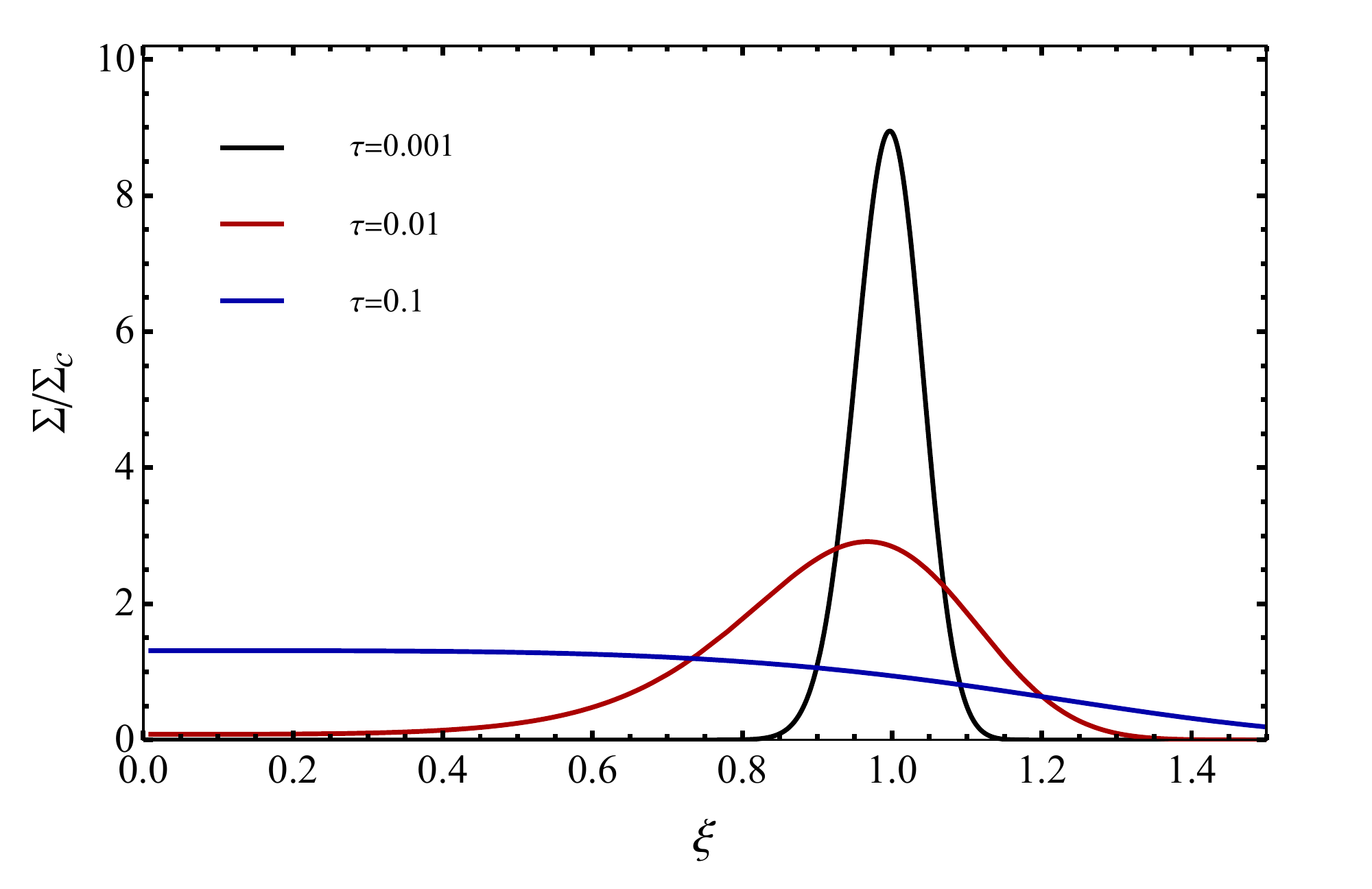}
\includegraphics[width=14cm]{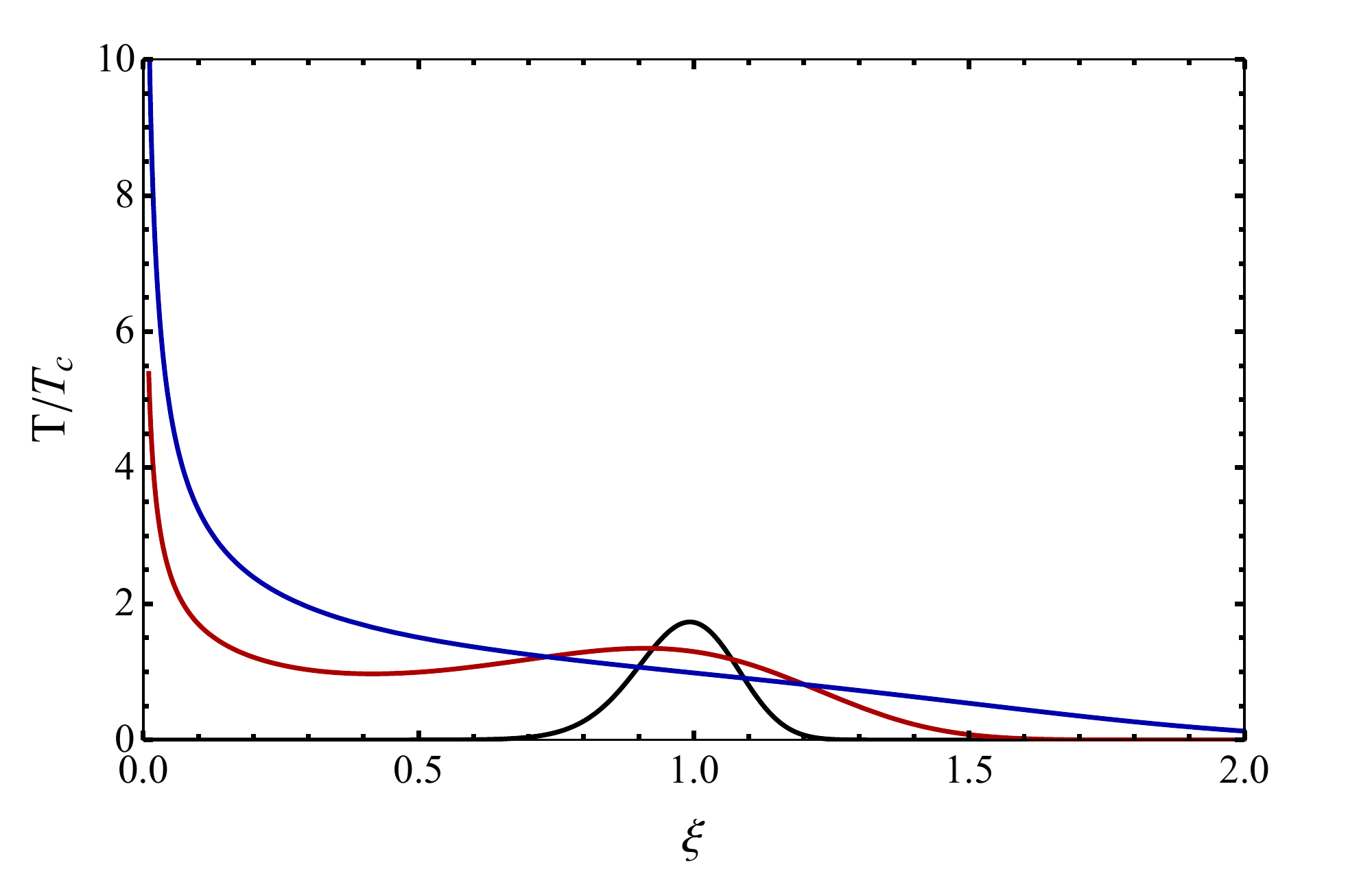}
\caption{
Evolution of the surface density profile ({\it top panel}) and the corresponding temperature evolution ({\it bottom panel}) as a function of the dimensionless disk radius expressed in units of the circularization radius ($\xi=\sqrt{r/r_{\rm c}}$) for a time-dependent accretion disk. The $\tau$ represents time normalized by $t_{\rm c}$ (the viscous time at $r_{\rm c}$, see equation 12).
In both panels, the different colors indicate the associated time $\tau$. The mass is initially distributed in a narrow ring around the circularization radius and it spreads and heats up (at the inner radii) due to conservation of angular momentum and viscous heating as time progresses.
}
  \label{fig:sdtemp}
\end{figure}

Adopting for the viscous stress tensor with a power-law index $\delta$
\begin{eqnarray}
\mathcal{T}_{r\phi}=\mathcal{T}_{r\phi,c}
\left(\frac{\Sigma}{\Sigma_{\rm c}}\right)
\left(\frac{r}{r_{\rm c}}\right)^{\delta},
\label{eq:shearst}
\end{eqnarray}
we can make equation (\ref{eq:diskevo}) dimensionless \citep{1998bhad.book.....K,2008bhad.book.....K}
\begin{eqnarray}
\frac{\partial \sigma}{\partial\tau}=\frac{1}{\xi^3}\frac{\partial^2}{\partial\xi^2}(\xi^{2\delta+4}\sigma),
\label{eq:diskevo2}
\end{eqnarray}
where $\sigma=\Sigma/\Sigma_{\rm c}$ with 
$\Sigma_{\rm c}$ being the surface density at the circularization radius $r_{\rm c}$, $\tau=t/t_{\rm c}$, and $\xi=(r/r_{\rm c})^{1/2}$. 
From equations (\ref{eq:tvis}) and (\ref{eq:shearst}), 
$t_{\rm vis}(r)\propto{r}^{1-2\delta}$ so that $\delta$ should be smaller than $1/2$ in order for the viscous 
timescale to be longer at larger radius. Because $\nu$ is only a power-law function of radius from equation~(\ref{eq:shearst}), we can analytically 
solve equation (\ref{eq:diskevo2}) by using the Green's function \citep{1974MNRAS.168..603L}.
The solution with the zero torque boundary is then given by
\begin{equation}
\sigma(\xi,\tau)
=
\mu\frac{\xi^{(2-9\mu)/2\mu}}{\tau}\exp\left[-\frac{\mu^2(\xi^{1/\mu}+1)}{\tau}\right]\mathcal{I}_{\mu}\left(\frac{2\mu^2\xi^{1/2\mu}}{\tau}\right),
\label{eq:sd}
\end{equation}
where $\mathcal{I}_\mu(z)$ is the modified Bessel function and $\mu=1/(1-2\delta)$.
We show the surface density evolution for the three different times in Figure~{\ref{fig:sdtemp}}, 
where $\mu=1/4$ (i.e., $\delta=-3/2$) is adopted.
The disk has initially a ring-like structure, which is shown in the black solid line, and subsequently 
evolves viscously. The red and blue solid lines show the surface densities at $\tau=0.01$ and $\tau=0.1$, respectively. The figure shows that the circularized ring-like most-bound debris formed at $\tau=0$ accretes onto the SMBH 
at $\tau=0.1$, leading to a delayed X-ray flare. The time it takes for the debris to be accreted by the SMBH can be evaluated using equation (\ref{eq:tc}):
\begin{equation}
t_{X}=\tau_{X}t_{\rm c}
\approx
1.9
\left(\frac{\tau_X}{0.1}\right)
\left(\frac{\alpha}{0.1}\right)^{-4/5}
\left(\frac{\zeta}{1/3}\right)^{-2/5}
\left(\frac{1+e_*}{2}\right)^{7/5}
\beta^{-17/5}
M_{\rm bh,6}^{-11/45}
m_{*,1}^{-19/15}
r_{*,1}^{2}
\,\,{\rm yr}\,,
\label{eq:tacc}
\end{equation}
where $\tau_{X}=t_X/t_{\rm c}$ and our solution indicates $\tau_X=0.1$ as a typical value of $\tau_X$.

The time delay between the optical and X-ray flares is then determined by the sum of $t_{\rm circ}$ and $t_{X}$ (equations \ref{eq:tcirc} and \ref{eq:tacc}) as
\begin{eqnarray}
\Delta t_{\rm O-X}
&=&t_{\rm circ} + t_{X}
\approx
8.3\,
\left(\frac{\eta}{0.1}\right)^{-1}
\beta^{-3}
M_{\rm bh,6}^{-7/6}\,m_{*,1}^{8/25}\,
\,\,{\rm yr}
\nonumber \\
&+&
1.6\,
\left(\frac{\tau_X}{0.1}\right)
\left(\frac{\alpha}{0.1}\right)^{-4/5}
\left(\frac{\zeta}{1/3}\right)^{-2/5}
\left(\frac{1+e_*}{2}\right)^{7/5}
\beta^{-17/5}
M_{\rm bh,6}^{-11/45}
m_{*,1}^{37/75}
\,\,{\rm yr}\,,
\label{eq:tox}
\end{eqnarray}
where we adopt the approximate formula of the stellar mass-radius relation: 
\begin{eqnarray}
r_{*,1}=0.93\,m_{*,1}^{0.88}, 
\label{eq:mr}
\end{eqnarray}
which is valid over the range of $0.15\,M_\odot\le{m_*}\le3\,M_\odot$  \citep{2020ApJ...904...99R}, incorporating the dependence on $r_{*,1}$ into that of $m_{*,1}$. We find that \dt\,depends strongly on the penetration factor $\beta$, while it depends less strongly on mass of the disrupted star. While $t_{\rm circ}$ becomes rapidly shorter with increasing  black hole mass, $t_{X}$ depends less strongly on the black hole mass. As a result, \dt is shorter the larger the black hole mass. By substituting equation~(\ref{eq:mr}) into equation (\ref{eq:prange}), the condition that $\dot{M}_{\rm c}<\dot{M}_{\rm Edd}$ is simplified to be
\begin{eqnarray}
\zeta\,\eta^{5/3}\,\beta^5\,m_{*,1}^{17/25}\,M_{\rm bh,6}^{23/18}\,
\lesssim7.6\times10^{-3}
\label{eq:prange1}
\end{eqnarray}
This inequality shows the parameter space where our model is applicable for given black hole and stellar masses. Figure~\ref{fig:prange} shows the dependence of $\dot{M}_{\rm c}/\dot{M}_{\rm Edd}$ on the penetration factor. 
Panel (a) represents our fiducial model with $M_{\rm bh,6}=1$ and $m_{*,1}=1$, while panel (b) shows 
the case for a lower stellar mass $m_{*,1}=0.15$. Panels (c) and (d) are for $M_{\rm bh,6}=0.1$, and also panels (e) and (f) are for $M_{\rm bh,6}=10$. The solid black, red, and blue lines denote the $(\eta,\zeta)=(0.1,1/3)$, $(\eta,\zeta)=(0.01,1/3)$, and  $(\eta,\zeta)=(0.1,1/12)$ cases, respectively. While the sub-Eddington accretion is dominant in the region below the horizontal dashed line, the super-Eddington accretion is dominant in the region above the horizontal dashed line.

Figure~\ref{fig:dt} shows the dependence of \dt on $\beta$ for different stellar and black hole mass and for given other parameters: $\tau_X=0.1$, $\alpha=0.1$, and $e_*=1$. Note that similar to the effect of changing $M_{\rm bh}$ or $m_{*}$,  
changing $\tau_X$ or $\alpha$ affects the height of each curve. The solid line represents \dt, whereas the dotted line represents $t_X$. Both \dt and $t_X$ decrease rapidly with increasing $\beta$ as seen in equations (\ref{eq:tacc}) and (\ref{eq:tox}). 
Since \dt is defined as the sum of the circularization timescale and $t_X$, the difference between the solid and dotted lines gives the circularization timescale. The circularization timescale accounts for a large fraction of \dt in a large part of the parameter space.
As seen in equation (\ref{eq:tox}), the circularization timescale $t_{\rm circ}$ is proportional to $M_{\rm bh}^{-7/6}$ whereas $t_X$ is proportional to $M_{\rm bh}^{-11/45}$ so that 
\dt is closer to $t_X$ with increasing black hole mass. In fact, as seen in panels (e) and (f), the difference is significantly smaller for $M_{\rm bh,6}=10$ case than for the other two cases of the $M_{\rm bh,6}=0.1$ and $M_{\rm bh,6}=1$. Also, $\eta$ is an important parameter for estimating \dt; a smaller $\eta$ results in a longer $t_{\rm circ}$ and by definition a longer \dt. Since $\eta$ does not affect $t_X$, the deviation between the solid and dotted lines becomes larger for a smaller $\eta$. In summary, 
$\Delta{t}_{\rm O-X}$ is dominated by ${t}_X$ for $M_{\rm bh,6}\gg1$ and a larger $\eta$, while $\Delta{t}_{\rm O-X}$ is dominated by ${t}_{\rm circ}$ for $M_{\rm bh,6}\ll1$ and a smaller $\eta$. The possible range for $\beta$ such that the accretion rate is sub-Eddington is seen in the corresponding panels of Figure~\ref{fig:prange}. We conclude that our model can explain an observed time delay of a year to dozens of years between the optical and X-ray flares.

%
\subsection{Spectra of a time-dependent disk}
%
Since the disk is effectively optically thick, the energy liberated by viscous heating is emitted through 
black body radiation. That is to say, $Q_{\rm rad}=Q_{\rm vis}$, where $Q_{\rm rad}=2\sigma{T^4}$ and $Q_{\rm vis}=-(3/2)\mathcal{T}_{r\phi}\Omega$.
We then obtain the blackbody temperature by substituting equation~(\ref{eq:shearst}) into $2\sigma{T^4}=-(3/2)\mathcal{T}_{r\phi}\Omega$,
\begin{eqnarray}
T(\xi,\tau)=T_{\rm c}
\left(\frac{\mathcal{T}_{r\phi}}{\mathcal{T}_{r\phi,{\rm c}}}\right)^{1/4}
\left(\frac{\Omega}{\Omega_{\rm c}}\right)^{1/4}
=
T_{\rm c}
\left(\frac{\Sigma}{\Sigma_{\rm c}}\right)^{1/4}
\left(\frac{r}{r_{\rm c}}\right)^{\delta/4-3/8}
=
T_{\rm c}
\sigma(\xi,\tau)^{1/4}\xi^{\delta/2-3/4}.
\label{eq:tempxi}
\end{eqnarray} 
The lower panel of Figure~\ref{fig:sdtemp} depicts the corresponding temperature evolution of the disk.

In the spectral range where the electron scattering opacity $\kappa_{\rm es}=0.4\,{\rm cm^2\,g^{-1}}$ dominates the free-free absorption opacity $\kappa_{\rm ff}=1.5\times10^{25}\rho{T}^{-7/2}(1-e^{-h\nu/kT})/(h\nu/k_{\rm b}T)^3\,{\rm cm^2\,g^{-1}}$, the emergent specific intensity, $I_\nu$, is modified from $B_\nu$ to be $\kappa(\nu,T)B_\nu$ \citep{1972A&A....17..226F,1973A&A....24..337S}, where $h$ is the Planck constant, $k$ is the Boltzmann constant, $B_\nu$ is the Planck function, and $\kappa(\nu,T)$ is a distortion factor for the outgoing radiation from the disk surface:
\begin{eqnarray}
\kappa(\nu,T)=\frac{2}{1+\sqrt{(\kappa_{\rm ff}+\kappa_{\rm es})\kappa_{\rm ff}^{-1}}}
\label{eq:kappa}
\end{eqnarray}
\citep{1979rpa..book.....R,2008bhad.book.....K}. 
Note that $\kappa(\nu,T)\approx1$ in the spectral range of $\kappa_{\rm es}\ll\kappa_{\rm ff}$ so that $I_\nu\approx{B}_\nu$. The flux density emerging from an accretion disk is expressed in the solid angle integral \citep{2002apa..book.....F,2008bhad.book.....K} as
\begin{eqnarray}
F_\nu
&=&
2\pi\frac{\cos{i}}{D^2}
\int_{R_{\rm in}}^{R_{\rm out}}\,I_{\nu}\,r\,dr
=
4\pi\frac{\cos{i}}{D^2}\frac{h}{c^2}\nu^3\int_{R_{\rm in}}^{R_{\rm out}}\frac{r\kappa(\nu,T)}{e^{h\nu/kT}-1}dr,
\label{eq:fnu}
\end{eqnarray}
where $i$ is the angle between the observer's line of sight and the normal to the disk plane, 
$D$ is the distance to the Earth, and $R_{\rm in}$ and $R_{\rm out}$ are the radii of the disk inner and outer edge, respectively. 
The spectral luminosity follows from equation (\ref{eq:fnu}) as
\begin{eqnarray}
L_\nu(\xi,\tau)=4\pi{D^2}\nu{F}_{\nu}=32\pi^2\cos{i}\frac{h\nu^4r_{\rm c}^2}{c^2}\int_{\xi_{\rm in}}^{\xi_{\rm out}}\frac{\xi^3\kappa(\nu,T)}{e^{h\nu/kT(\xi,\tau)}-1}d\xi,
\label{eq:lnu_xi}
\end{eqnarray}
where $\xi_{\rm in}=R_{\rm in}/r_{\rm c}$, $\xi_{\rm out}=R_{\rm out}/r_{\rm c}$, $\rho_{c}=\Sigma_{\rm c}/(2H_{\rm c})$ is the mass density at $r_{\rm c}$.
For the purpose of comparing equation (\ref{eq:lnu_xi}) with the blackbody spectral luminosity emitted from the standard disk, 
setting $x\equiv{h\nu}/kT$ together with $T=T_{\rm in}(r/R_{\rm in})^{-p}$, the spectral luminosity is estimated to be 
\begin{eqnarray}
L_\nu
&&
=
16\pi^2
\cos{i}
\frac{1}{p}
\frac{h\nu^4R_{\rm in}^2}{c^2}
\left(\frac{kT_{\rm in}}{h\nu}\right)^{2/p}
P(\nu;x_{\rm in},x_{\rm out})
\nonumber \\
&&
=4.0\times10^{44}\,
\left(\frac{3/4}{p}\right)
\left(\frac{k_{\rm b}T_{\rm in}}{h\nu_{\rm in}}\right)^{2/p}
\left(\frac{T}{T_{\rm in}}\right)^{2/p}
\left(\frac{\nu}{\nu_{\rm in}}\right)^{4-2/p}
\left[\frac{P(\nu;x_{\rm in},x_{\rm out})}{P(\nu;0,\infty)}\right]
\,{\rm erg/s}\,,
\label{eq:lnu}
\end{eqnarray}
where $p=3/4$ is the standard disk value, $x_{\rm in}=h\nu/kT_{\rm in}$, $x_{\rm out}=x_{\rm in}(R_{\rm out}/R_{\rm in})^p$, $R_{\rm in}=3r_{\rm S}\approx8.9\times10^{11}\,(M_{\rm  bh}/10^6M_\odot)\,{\rm cm}$, 
$T_{\rm in}=\epsilon_{\rm obs}/k_{\rm b}\approx1.2\times10^6\,(\epsilon_{\rm obs}/0.1\,{\rm keV})\,{\rm K}$, 
and $\nu_{\rm in}=k_{\rm b}T_{\rm in}/h\approx2.4\times10^{16}\,(\epsilon_{\rm obs}/0.1\,{\rm keV})\,{\rm Hz}$ are 
adopted, and $P(\nu;x_{\rm in},x_{\rm out})\equiv\int_{x_{\rm in}}^{x_{\rm out}}x^{2/p-1}(e^{x}-1)^{-1}dx$ and $P(\nu;0,\infty)\approx1.93$. We note that $R_{\rm in}$ changes as a function of the black hole spin (\citealt{1972ApJ...178..347B}).

Figure~\ref{fig:lumi} shows the spectral luminosity at respective time steps: $\tau=0.001$, $\tau=0.01$, and $\tau=0.1$ for two different inclination angles. The blue solid line shows the spectral luminosity of our time-dependent model at $\tau_X=0.1$, whereas the green dashed line shows the spectral luminosity of a standard, i.e., steady-state, disk model corresponding to our mode at $\tau_X=0.1$.
It can be seen from Figure~\ref{fig:lumi} that there is no large difference in the spectral energy distribution between the $\tau_X=0.1$ disk and the standard disk save for the slightly larger luminosity for the standard disk at the low energy side. The magenta solid line represents the spectral luminosity calculated based on the multi-color modified blackbody spectrum of our model at $\tau_X=0.1$. The multi-color modified blackbody disk luminosity is significantly lower than the multi-color blackbody disk luminosity at the high-energy side, whereas it is in good agreement with the multi-color blackbody disk luminosity at the low-energy side. This is because, as seen in equation~(\ref{eq:kappa}), $I_\nu$ is smaller than $B_{\nu}$ for $\kappa_{\rm ff}\ll\kappa_{\rm es}$ in the high temperature region of the disk, whereas $I_\nu\approx{B_\nu}$ for $\kappa_{\rm ff}\gg\kappa_{\rm es}$ in the low temperature region. Also, it is clear that the optical luminosity coming from the disk at early times is very low. This suggests the origin of the optical emission should be sought elsewhere, and the shock-powered optical luminosity caused by the debris self-interaction is a good contender model. In contrast, at late times, after a fraction of the viscous timescale has passed, the soft X-ray luminosity is a small fraction of the Eddington luminosity if the line of sight corresponds to the disk mid-plane, i.e., $i\approx90^\circ$ (edge-on view). This can explain the late-time X-ray luminosity observed in several optically discovered TDEs (J20). On the other hand, the X-ray luminosity gets close to the Eddington luminosity in the $i=0^\circ$ (pole-on) case.

This is also consistent with what \cite{2018ApJ...859L..20D} proposed before, i.e., the soft-X-ray luminosity becomes higher (lower) when the viewing angle, $i$, decreases (increases). According to the general relativistic radiation magnetohydrodynamic simulations of \cite{2018ApJ...859L..20D}, the lower the viewing angle is, the more dominant the effect of the hotter temperature side of the blackbody radiation on the spectral distribution becomes, and the whole distribution shifts to the higher energy side, resulting in a larger X-ray luminosity with a lower viewing angle. In our model, on the other hand, since the spectral distribution is simply proportional to $\cos\theta$ as seen in equations~(\ref{eq:lnu_xi}) and (\ref{eq:lnu}), the X-ray luminosity becomes larger when the viewing angle is lower. The properties of the small sample of sources studied in J20, both their non-detection as well as the relatively low X-ray luminosities are not inconsistent with this picture because the X-ray luminosity depends strongly on several other parameters, e.g., $\beta$, $m_{*}$, and $r_{*}$, and the values in the TDEs studied in J20 can depart from the fiducial values we adopted here. 

%
\section{Summary and Discussion}
\label{sec:dis}
%

We have studied the black-body emission from a 1D time-dependent accretion disk formed after debris circularization 
to explain the significant time difference between the optical and and subsequent X-ray detection in several optically selected TDEs. We have solved equation (\ref{eq:diskevo}) using the separation of variables method and the Green's function. \cite{1990ApJ...351...38C} solved the same diffusion equation as equation (\ref{eq:diskevo}) using a self-similar solution method and then found that the light curve deviates from $t^{-5/3}$ at late times. Both methods are good for calculating the secular, long-term evolution of the disk. In particular, our model is suitable for the so-called initial value problem, i.e., how the disk accretes diffusively from the initial ring-like structure to the SMBH. However, since our model makes the simplifying assumption that the viscous torque is a power-law function of radius and surface density as in equation (\ref{eq:shearst}), the general behavior needs to be checked using numerical calculations. Also, if the disk is thermally unstable, we need to solve the disk evolution and structure numerically with the energy equation (e.g., \citealt{1991PASJ...43..147H}). These cases and their influence on \dt~will be investigated in future work.

\begin{figure}[!htbp]
\centering
\resizebox{8.5cm}{!}{\includegraphics{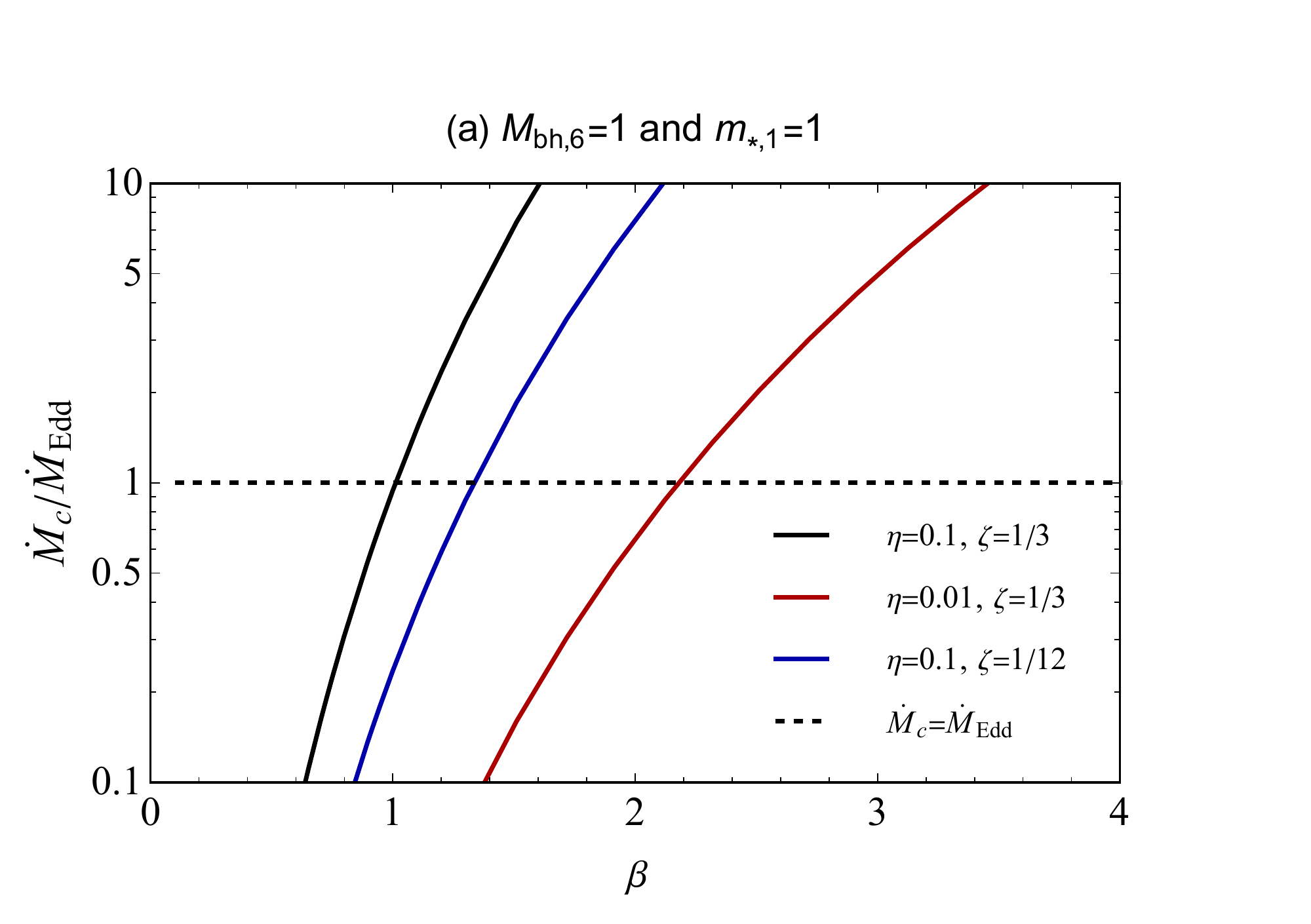}}
\resizebox{8.5cm}{!}{\includegraphics{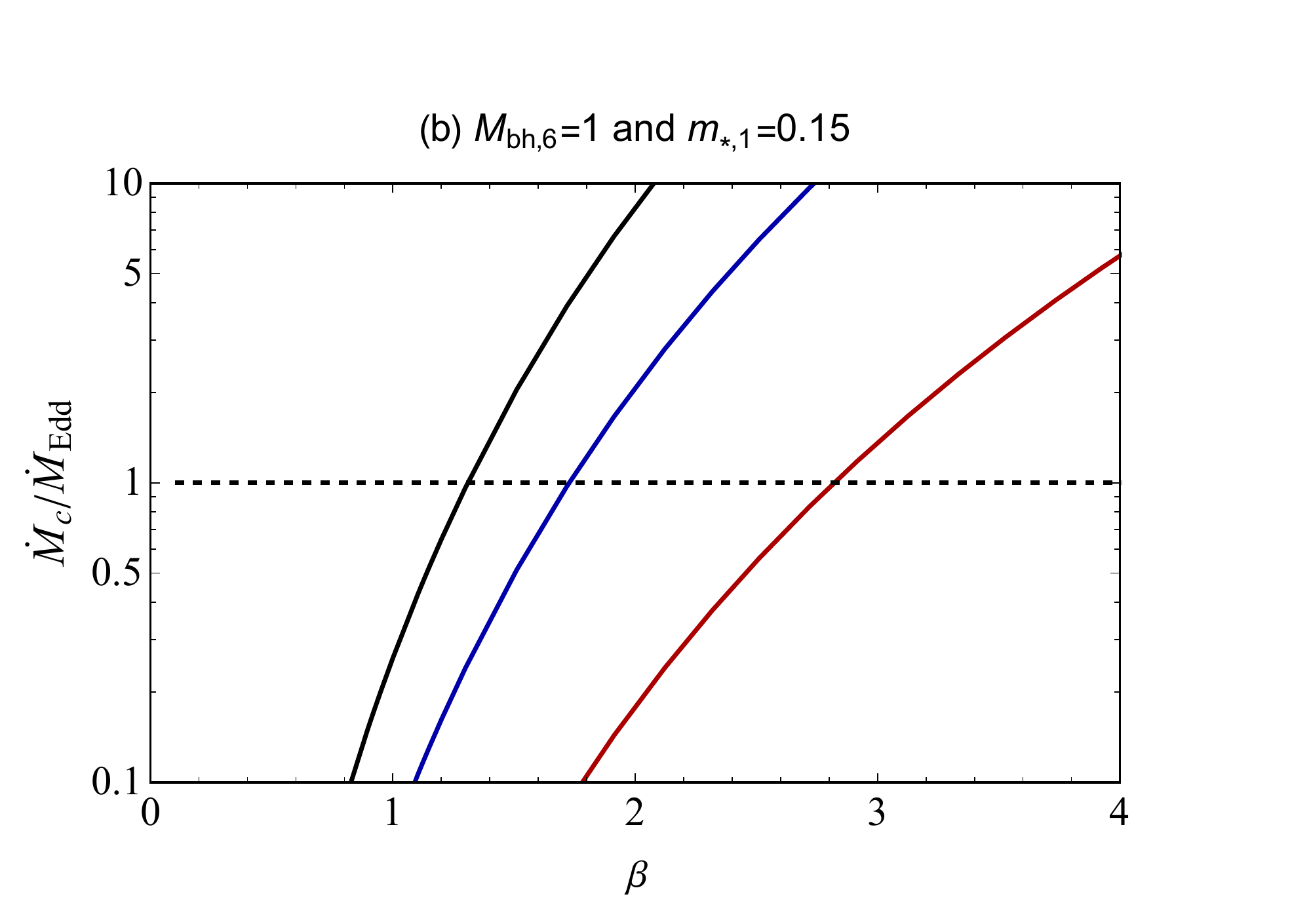}}\\
\resizebox{8.5cm}{!}{\includegraphics{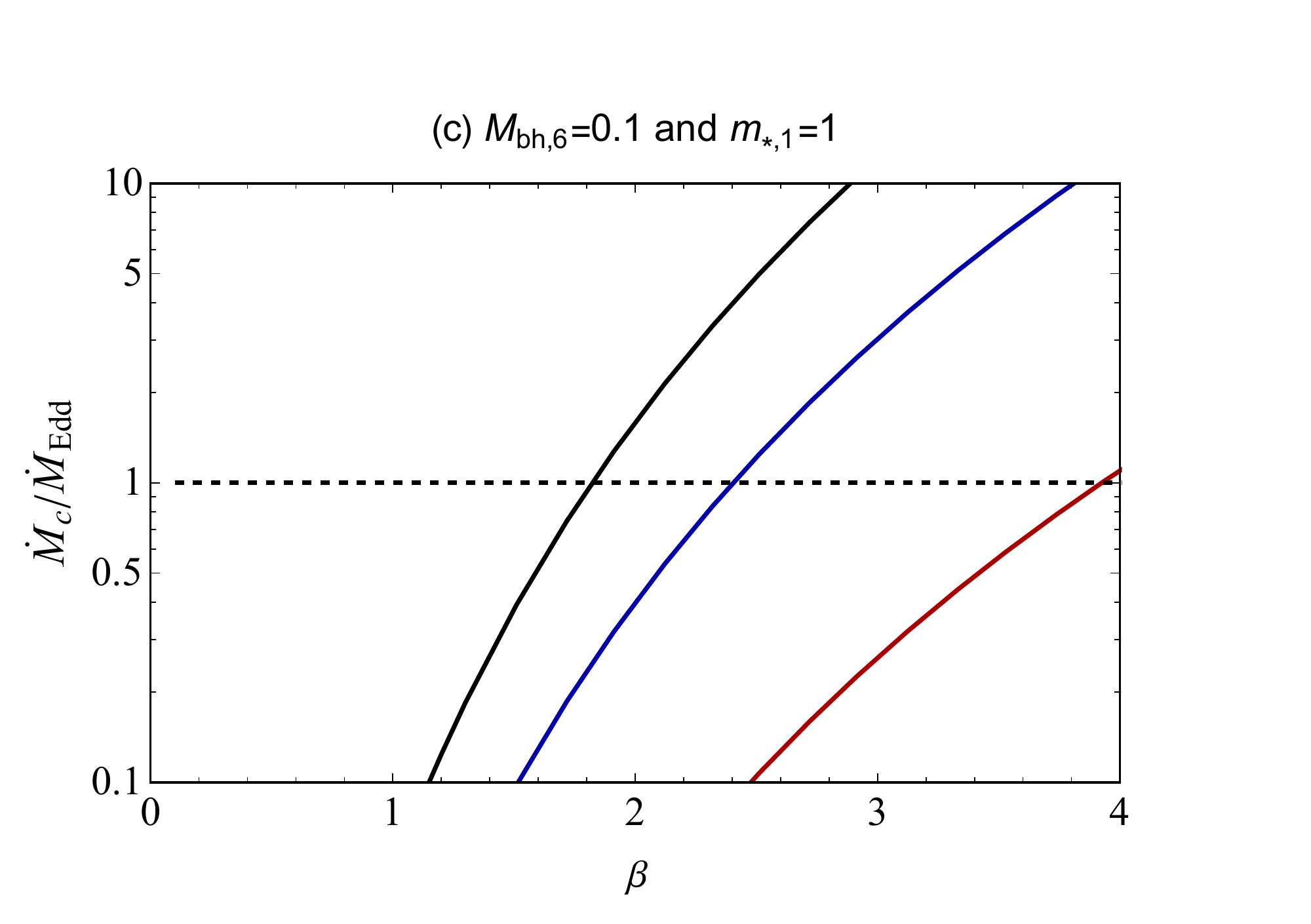}}
\resizebox{8.5cm}{!}{\includegraphics{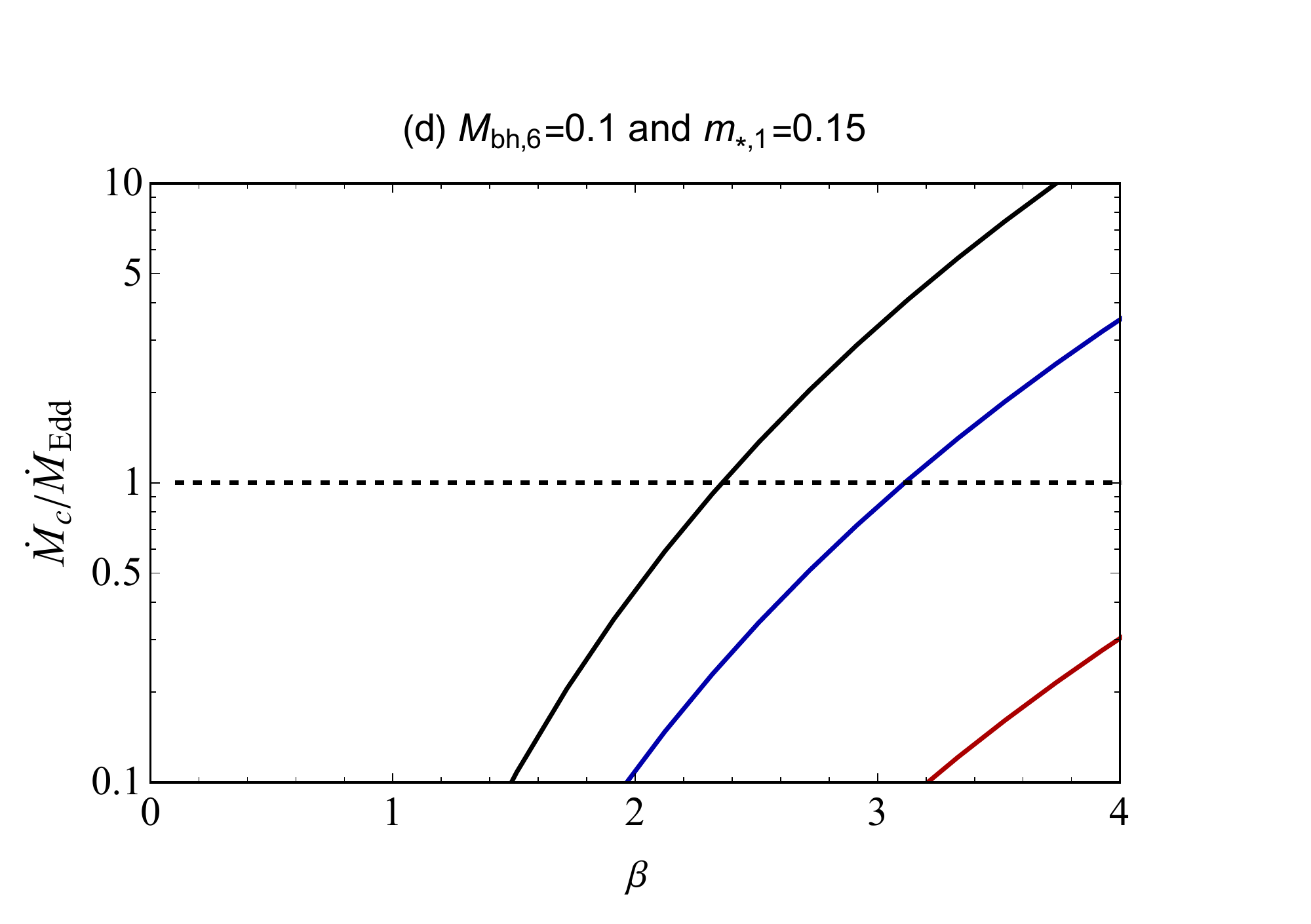}}\\
\resizebox{8.5cm}{!}{\includegraphics{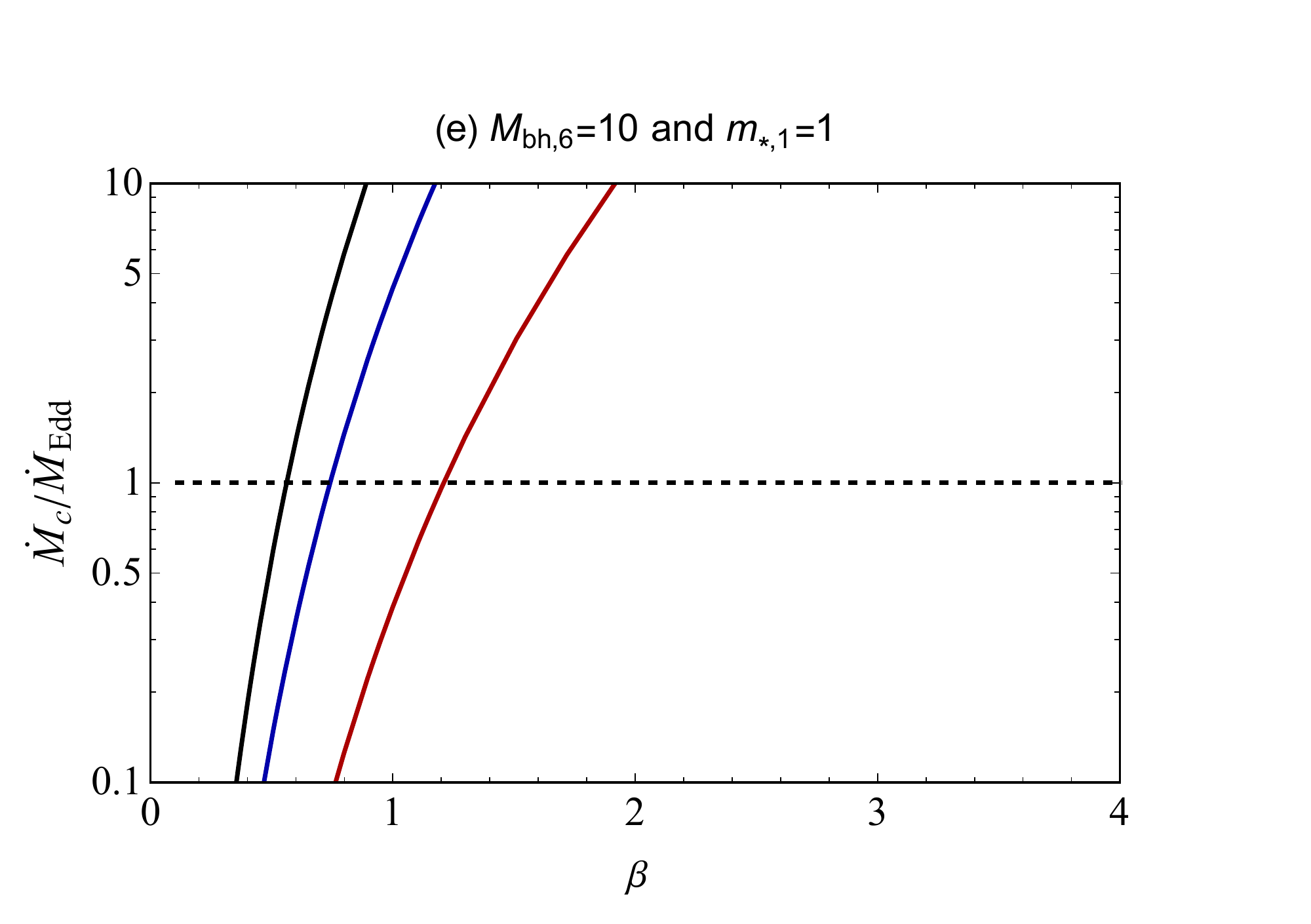}}
\resizebox{8.5cm}{!}{\includegraphics{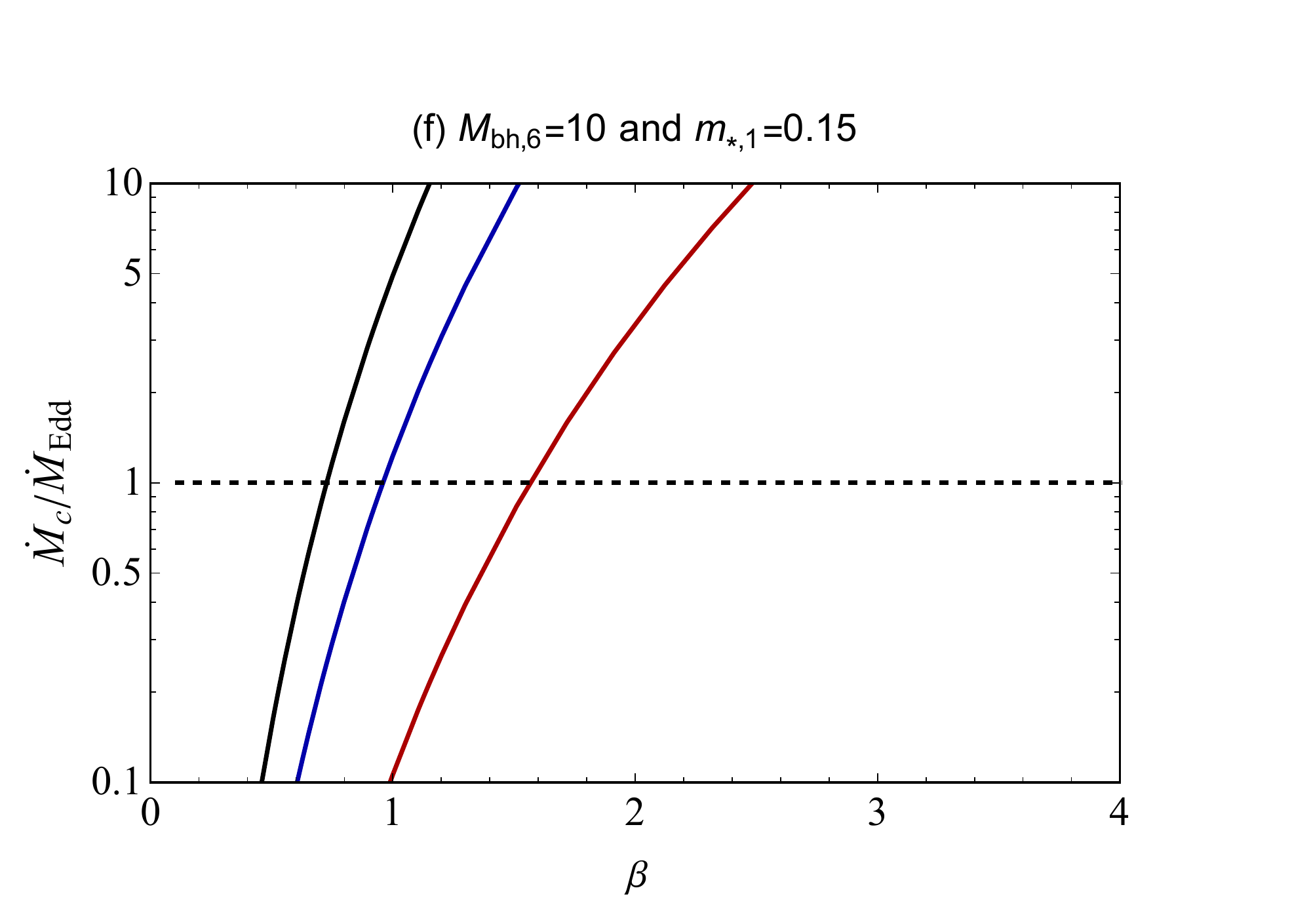}}
\caption{
Parameter space of sub- and super-Eddington accretion for given black hole and stellar mass (see also equation~\ref{eq:prange1}). Each panel shows the dependence of $\dot{M}_{\rm c}/\dot{M}_{\rm Edd}$ on the penetration factor. Panel (a) represents the fiducial model with $M_{\rm bh,6}=1$ and $m_{*,1}=1$, while panel (b) shows the case for the disruption of a lower  stellar mass $m_{*,1}=0.15$ star. Panels (c) and (d) are for $M_{\rm bh,6}=0.1$, and  panels (e) and (f) are for $M_{\rm bh,6}=10$. The solid black, red, and blue lines denote the $(\eta,\zeta)=(0.1,1/3)$, $(\eta,\zeta)=(0.01,1/3)$, and  $(\eta,\zeta)=(0.1,1/12)$ case, respectively.
}
\label{fig:prange}
\end{figure}

\begin{figure}[!htbp]
\centering
\resizebox{8.5cm}{!}{\includegraphics{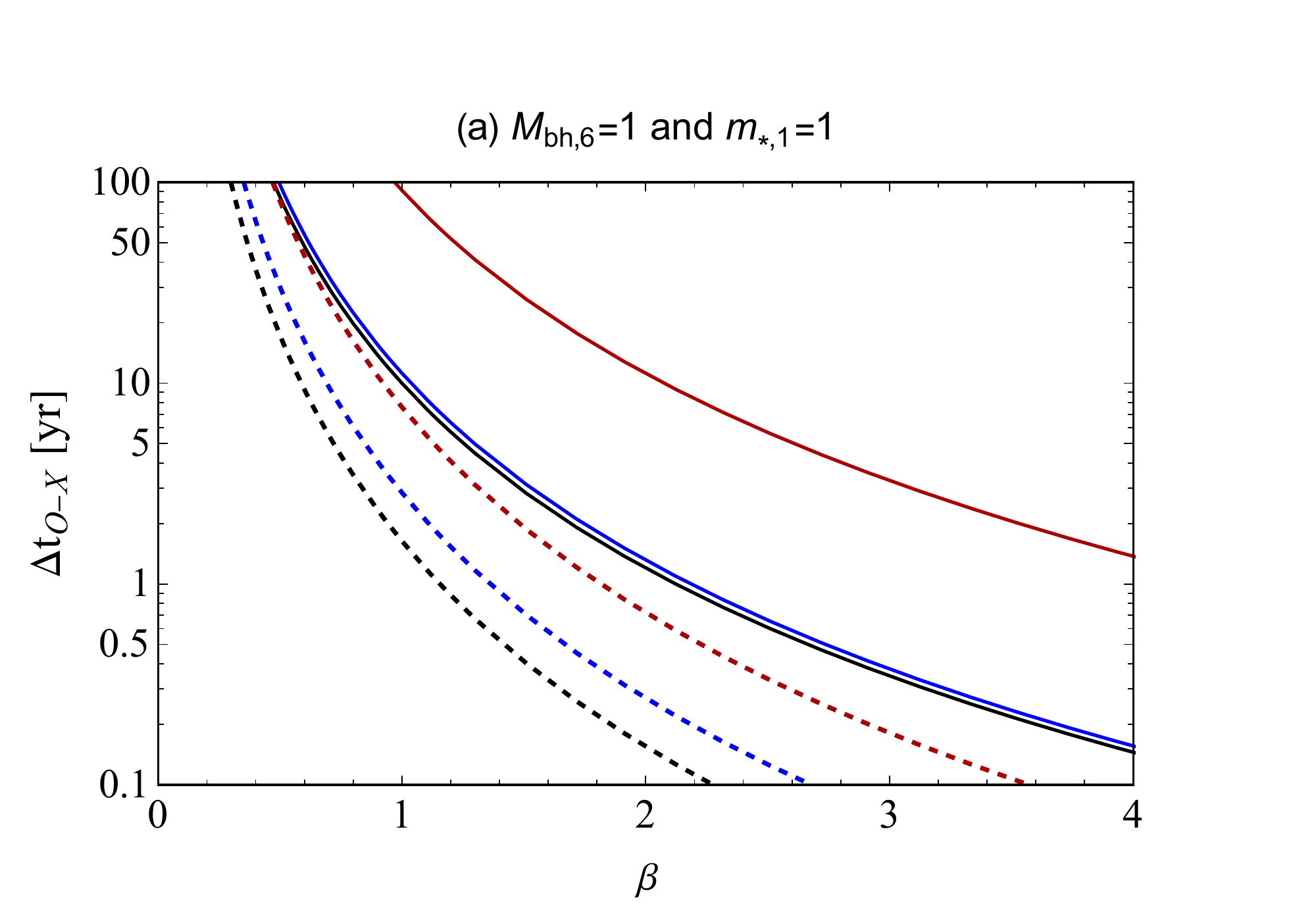}}
\resizebox{8.5cm}{!}{\includegraphics{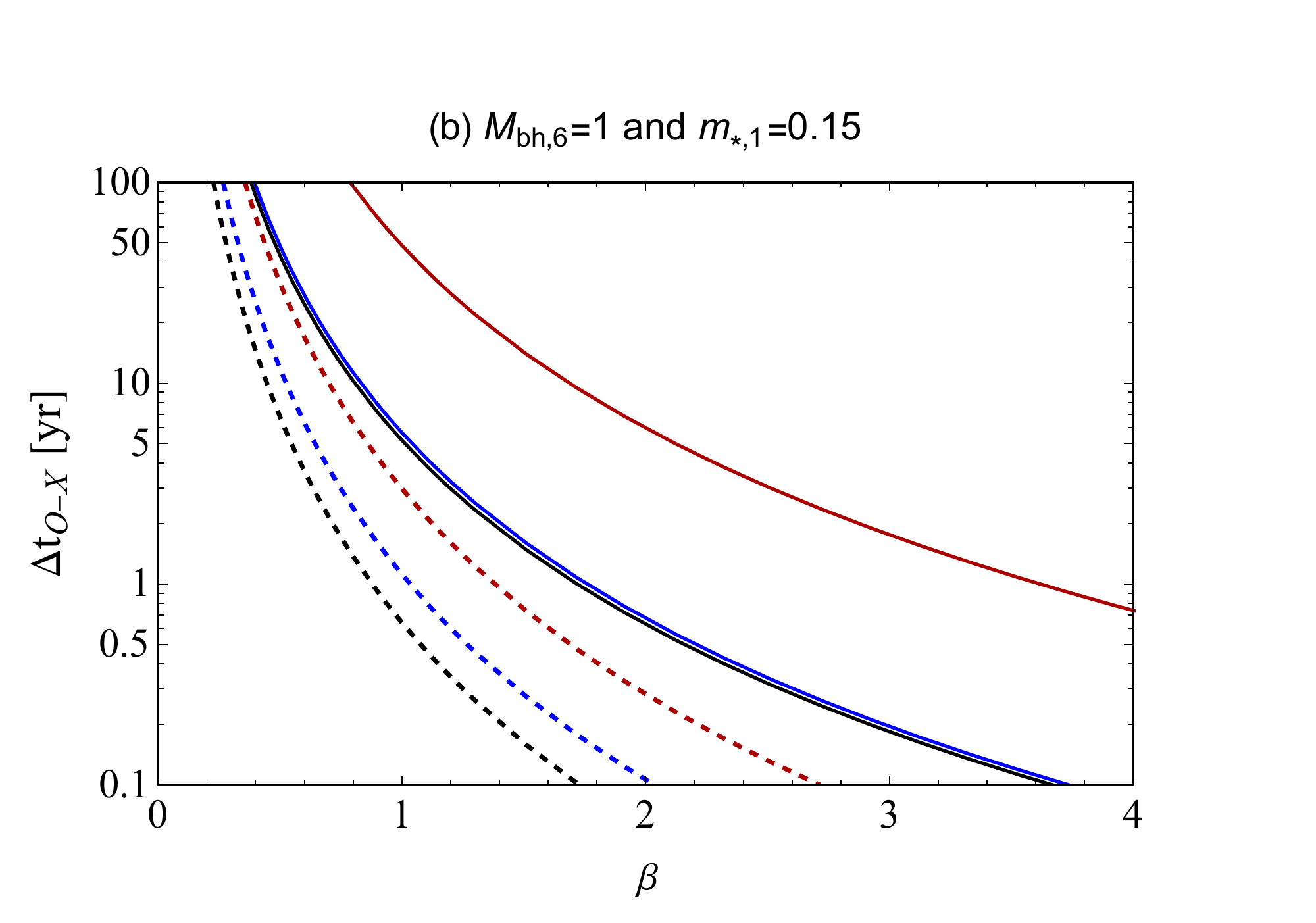}}\\
\resizebox{8.5cm}{!}{\includegraphics{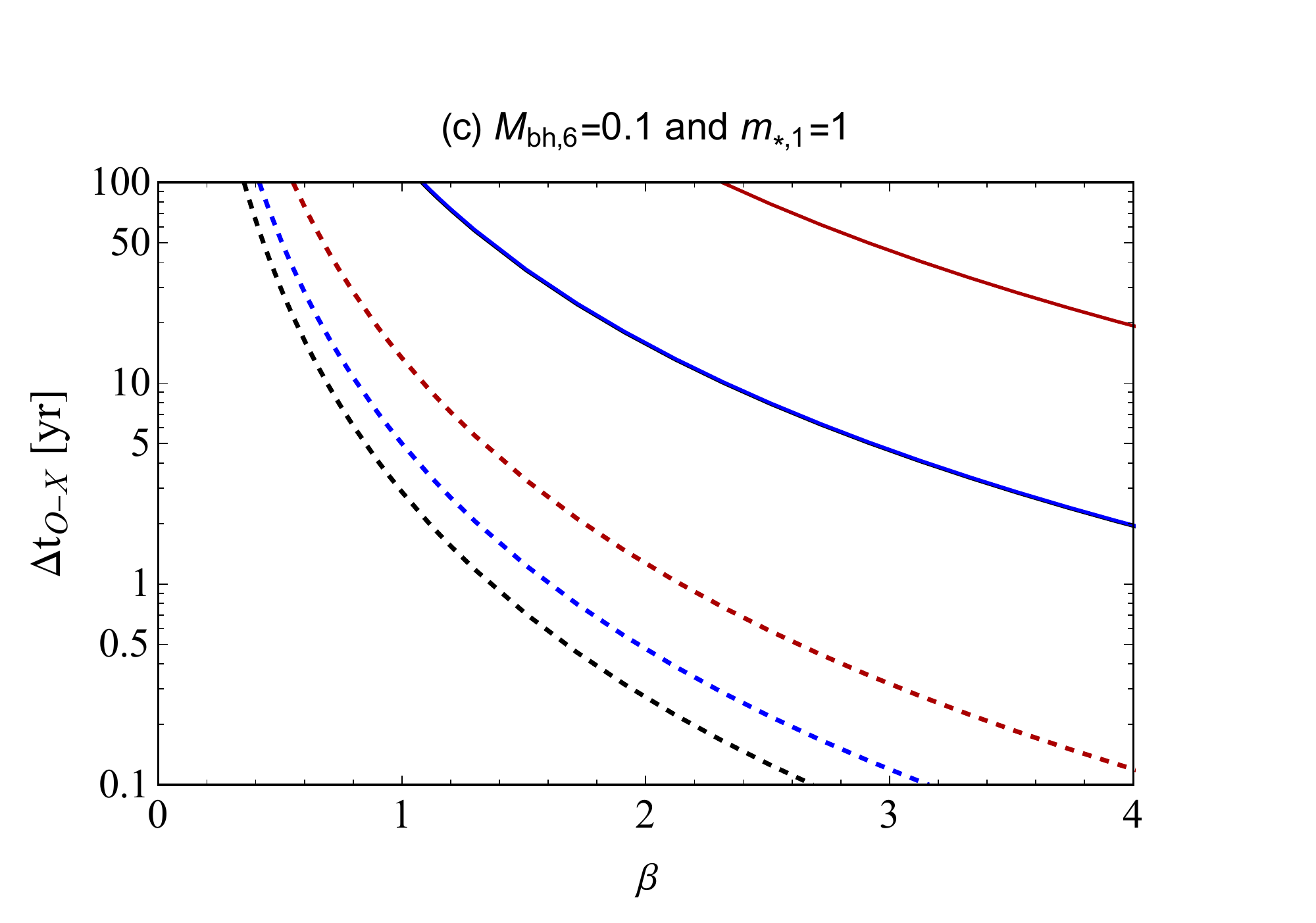}}
\resizebox{8.5cm}{!}{\includegraphics{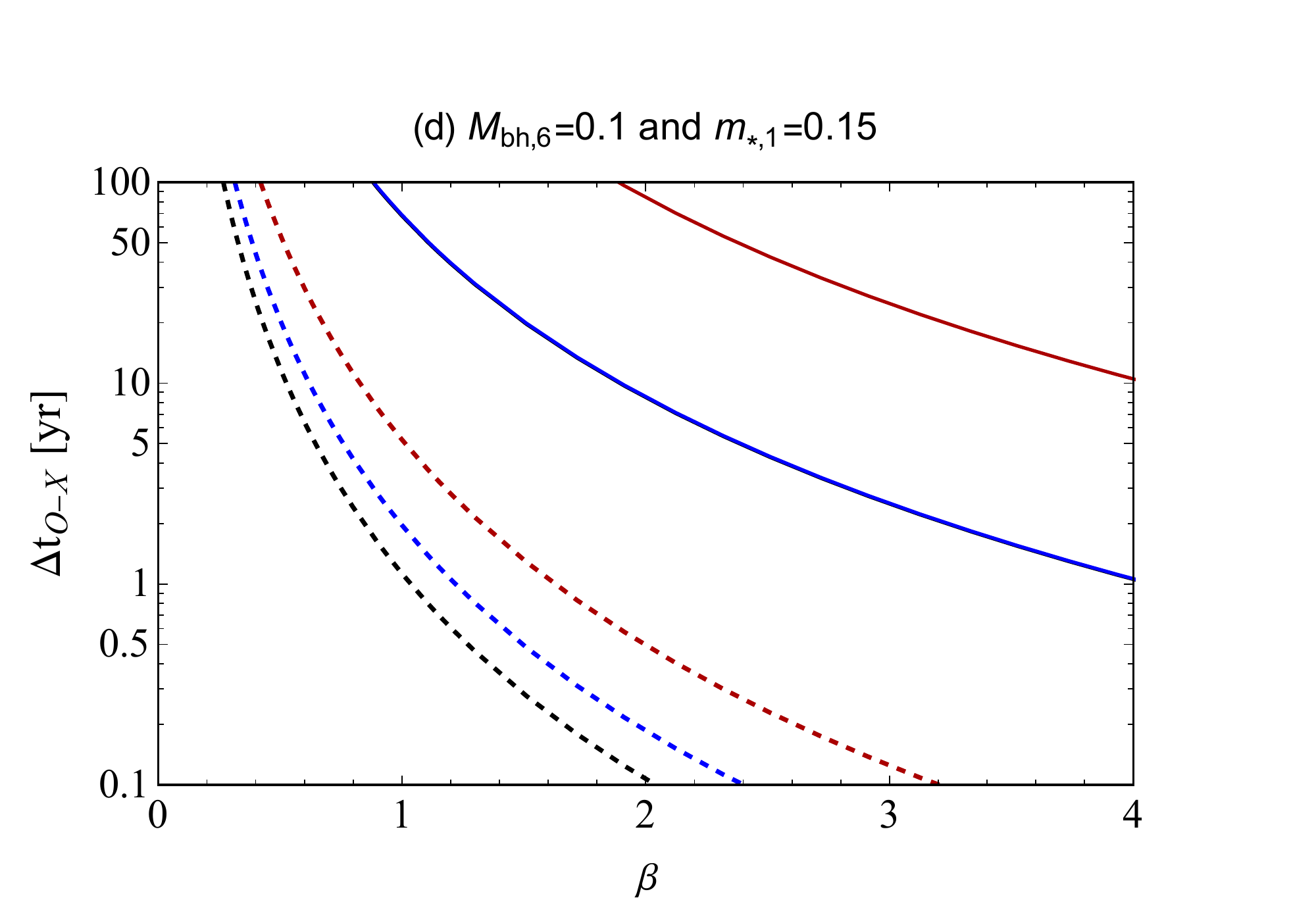}}\\
\resizebox{8.5cm}{!}{\includegraphics{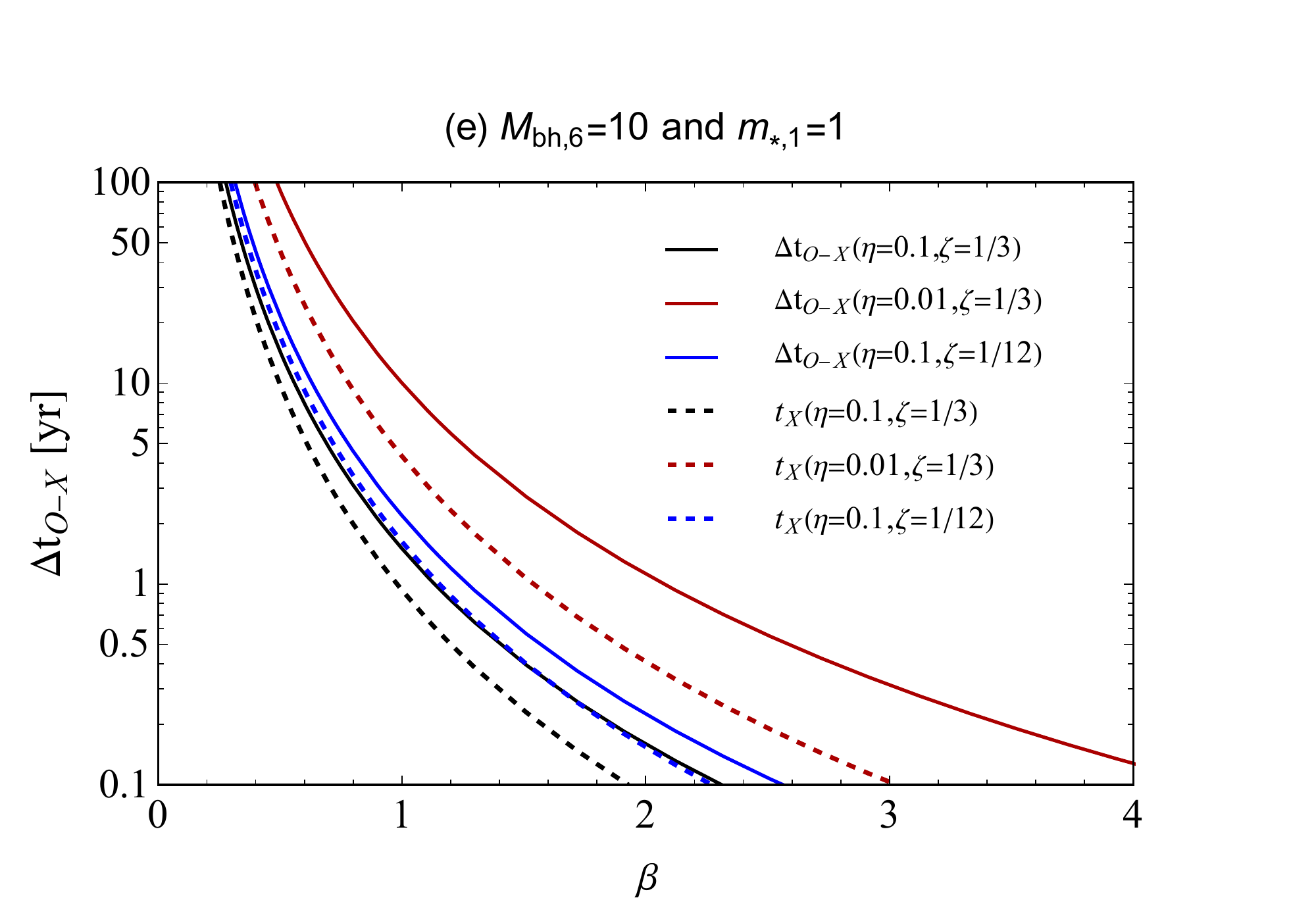}}
\resizebox{8.5cm}{!}{\includegraphics{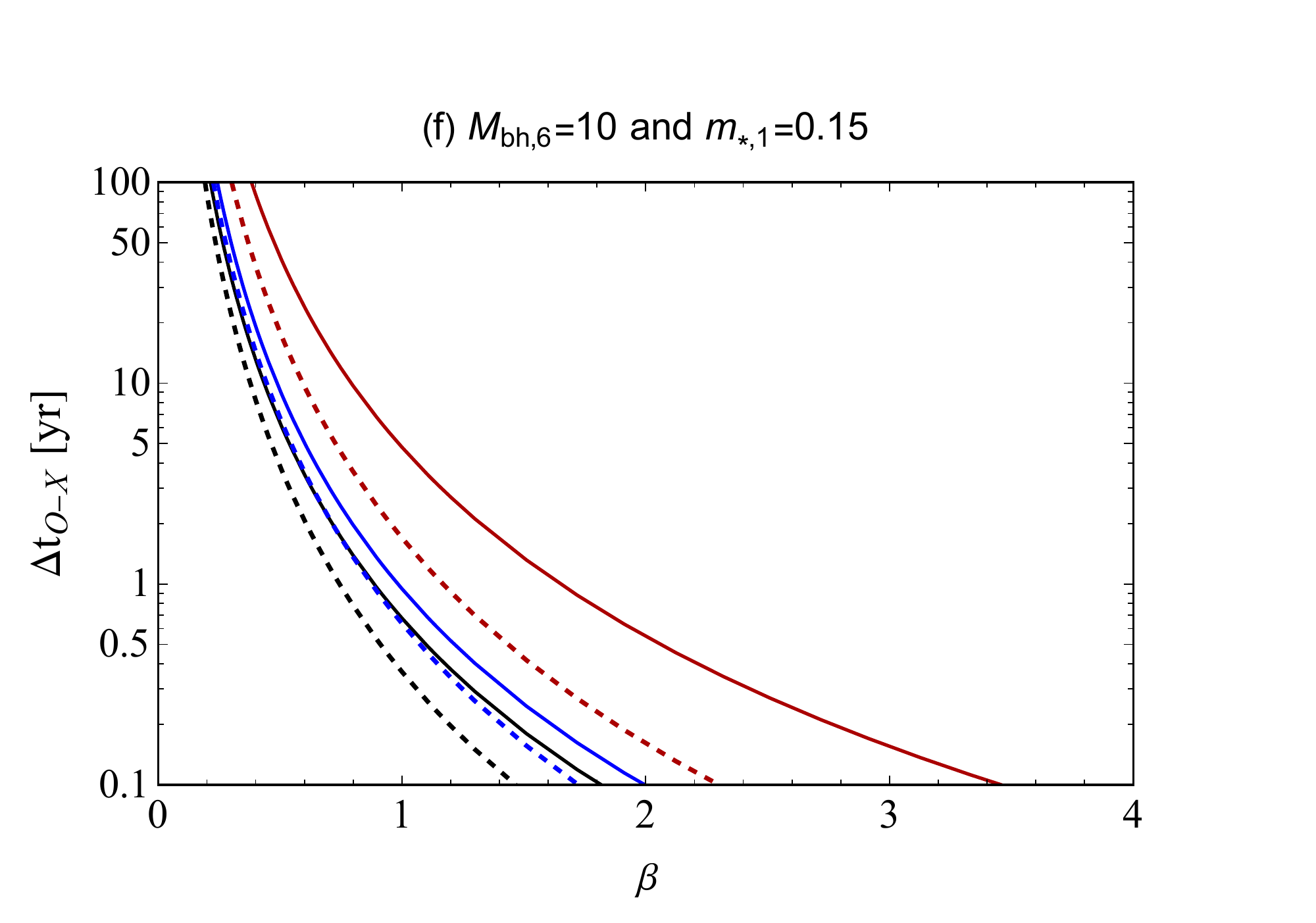}}
\caption{
Dependence of \dt on the penetration factor for given black hole and stellar mass (the other fiducial parameters and their values are $\tau_X=0.1$, $\alpha=0.1$, and $e_*=1$, see also equation \ref{eq:tox}). Panels (a,c,e) and (b,d,f) show the $m_{*,1}=1$ and $m_{*,1}=0.15$ cases, respectively. In each panel, the solid black, red, and blue lines denote the $(\eta,\zeta)=(0.1,1/3)$, $(\eta,\zeta)=(0.01,1/3)$, and $(\eta,\zeta)=(0.1,1/12)$ cases, respectively.
Each panel shows the dependence of \dt on the different black hole mass and the different stellar mass. Note that this mass dependence format of each panel is the same as that of Figure~\ref{fig:prange}.
}
\label{fig:dt}
\end{figure}

\begin{figure}[!htbp]
\centering
\includegraphics[width=11.5cm]{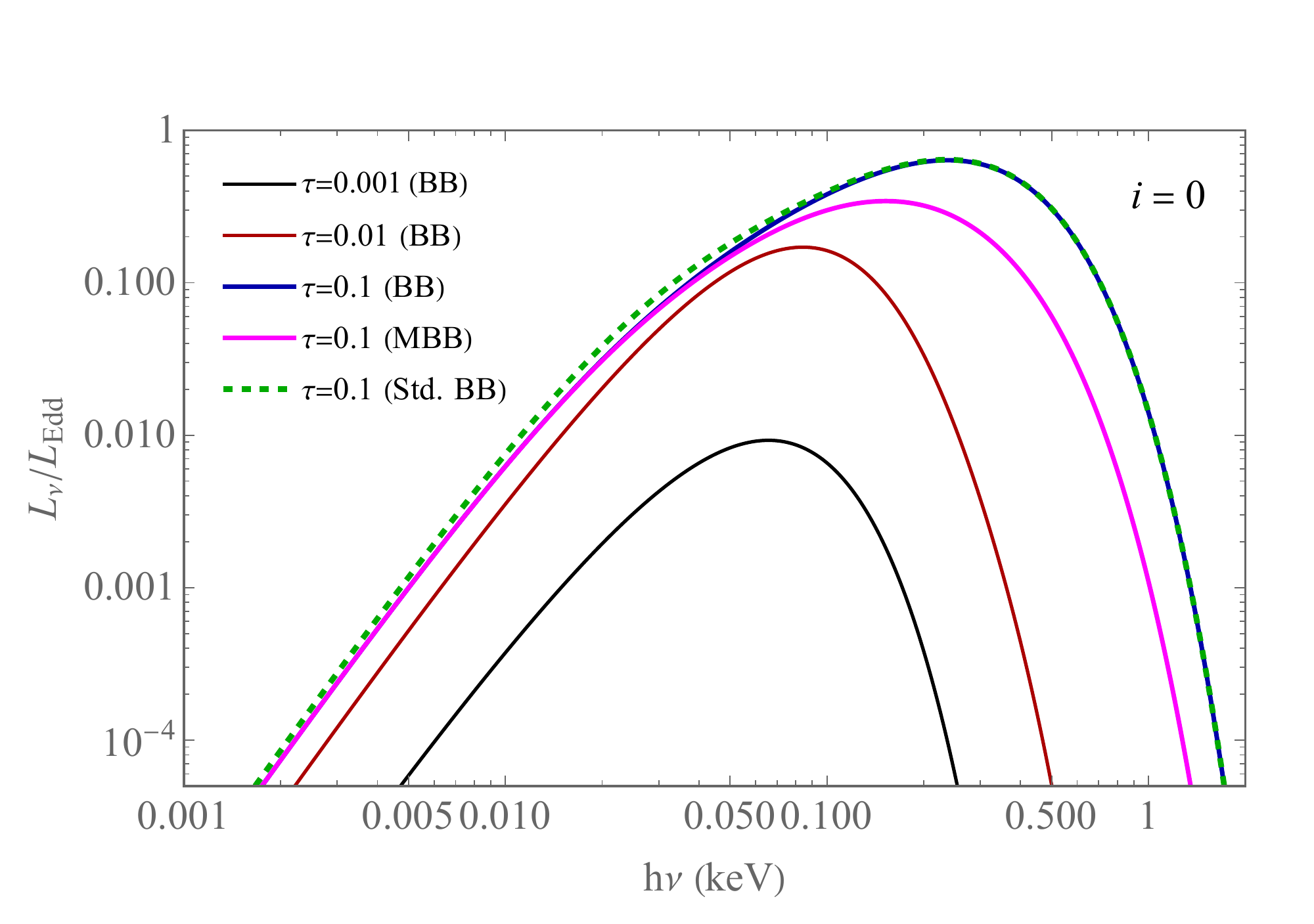}
\includegraphics[width=11.5cm]{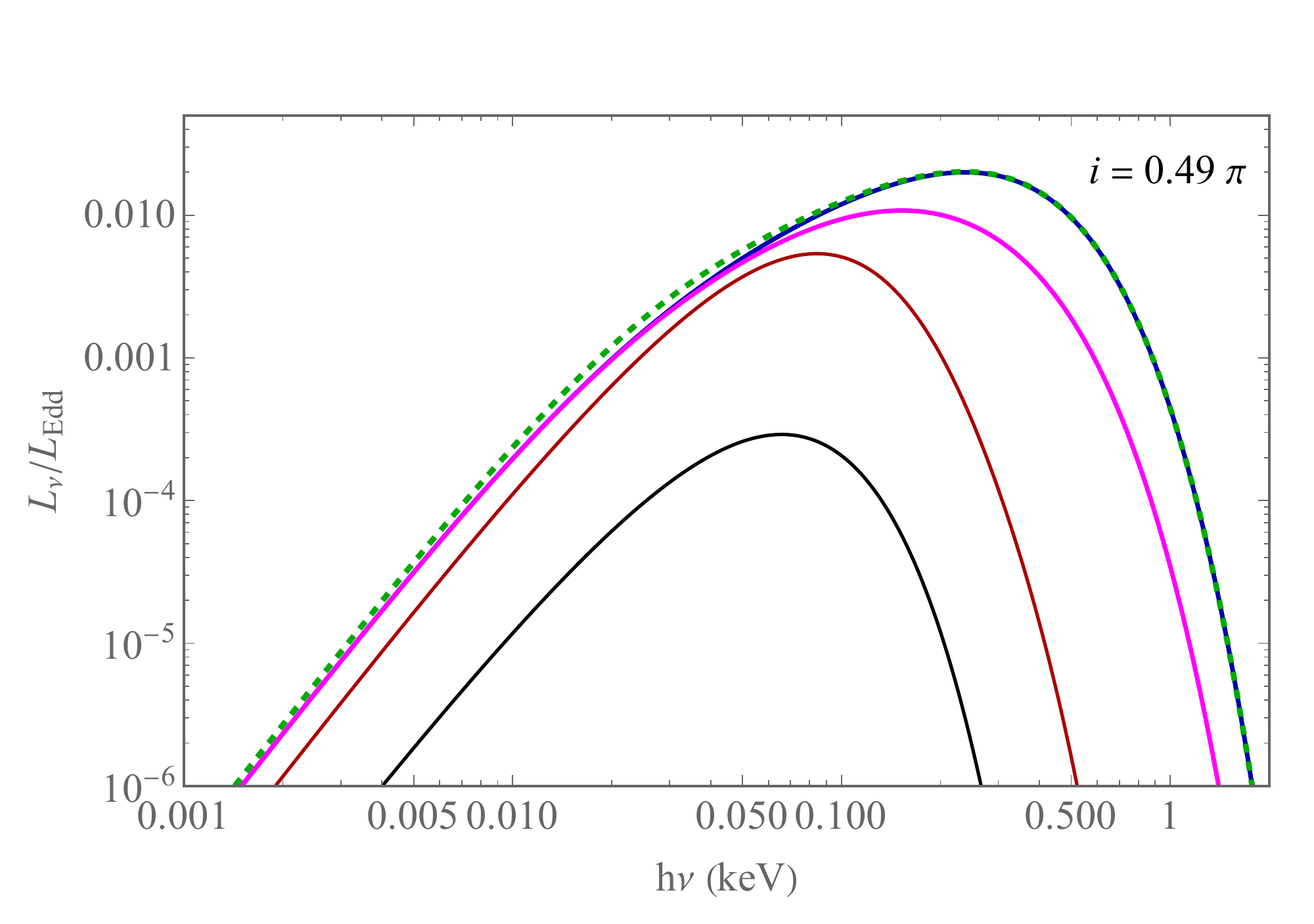}
\caption{
Evolution of the normalized spectral luminosity of the time-dependent disk corresponding to the two panels in Figure 1. In both panels, the black, red, and blue solid lines represent the disk spectral luminosities at $\tau=0.001$, $\tau=0.01$, and $\tau=0.1$ respectively computed using equation (\ref{eq:lnu_xi}) with $\kappa(\nu,T)=1$, while the dashed green line shows the spectral luminosity computed using equation (\ref{eq:lnu}) with the radial temperature profile of the standard disk. The magenta solid line represents the spectral luminosity of the modified blackbody disk spectrum at $\tau=0.1$. Here, the outgoing emission deviates from a blackbody disk spectrum due to electron scattering dominating the opacity, i.e., $\kappa(\nu,T)\neq1$. While the upper panel shows the $i=0$ (pole-on) case, the lower panel shows the $i\approx90^\circ$ (edge-on) case. Note the different values on the y-axis for the two panels.
}
\label{fig:lumi}
\end{figure}

According to the shock powered scenario, the early-time optical emission is caused by energy dissipation from the shock formed due to the debris stream self-interaction at the orbital semi-major axis of the most tightly bound debris. The debris's orbital energy is subsequently dissipated such that the most tightly bound debris is evolving towards circularization. The circularized disk viscously evolves as seen in the {\it top} panel (a) of Figure 1. The disk initially has a ring-like structure around the circularization radius. In our model, material of the disk accretes onto the SMBH by viscous diffusion for one-tenth of the viscous timescale. As seen in equation~(\ref{eq:tox}), the time delay between the optical and X-ray flares is the sum of the circularization timescale and the accretion timescale so that the viscous accretion is one of the two essential factors to cause the delayed X-ray emission. This delay amounts to $~\sim2$ years for typical parameters after the peak of optical emission in the optical/UV TDEs if the mass fallback rate is sub-Eddington (our fiducial parameters are a $10^6$~M$_\odot$ black hole, a $1~{\rm M}_{\odot}$ star, $\beta=1$, $e_*=1$, $\zeta=1/3$, and $\alpha=0.1$). We find that the viscous timescale is most sensitive to the penetration factor, $\beta$, whereas the viscous timescale is relatively insensitive to the black hole mass. Note that although the second-most sensitive parameters are stellar mass and radius, the effect of the two parameters on $t_{\rm X}$ cancels each other out by the mass-radius relation for the main-sequence star, resulting in being insensitive for the stellar mass. On the other hand, the circularization timescale also strongly depends on $\beta$ and is sensitive to the circularization efficiency and the black hole mass, whereas it is insensitive to the stellar mass. The delay by the debris circularization amounts $\sim8$ years for the fiducial parameters, where we adopt $\eta=0.1$ as the fiducial circularization efficiency. The resultant time delay \dt significantly decreases with increasing penetration factor. This strong dependence of $\beta$ on \dt is seen in Figure~\ref{fig:dt} with the dependence of the other parameters. Note that these parameters should satisfy the condition for the sub-Eddington accretion $\zeta\,\eta^{5/3}\,\beta^5\,m_{*,1}^{17/25}\,M_{\rm bh,6}^{23/18}\,\lesssim7.6\times10^{-3}$ (see equation~\ref{eq:prange1} and Figure~\ref{fig:prange}). The fiducial \dt is longer than $\sim10\,{\rm years}$ for $\beta\lesssim1$, while it is shorter than $1\,{\rm year}$ for $\beta\gtrsim2.2$. This suggests that the observed long-term \dt  ($4\sim10\,{\rm years}$) for several TDEs involves a star that is tidally disrupted at $\beta\gtrsim1$, while the observed short-term \dt ($\sim1\,{\rm year}$) for TDEs correspond to relatively high penetration factor disruptions.

Next, let us consider what happens to the optical-X-ray time delay if mass accretes onto the SMBH at a super-Eddington rate, $\dot{M}_{\rm c}>\dot{M}_{\rm Edd}$ (see equation \ref{eq:mdotc}), i.e., when $\zeta\,\eta^{5/3}\,\beta^5\,m_{*,1}^{17/25}\,M_{\rm bh,6}^{23/18}\,>7.6\times10^{-3}$. 
The viscous timescale is $10^4$ times as short for the geometrically thick ($H/r\sim1$) compared to the geometrically thin disk case (see equation~\ref{eq:tvis}) so that $t_X<t_{\rm vis}\ll{t}_{\rm circ}$. Therefore, $\Delta{t}_{\rm O-X}=t_X+t_{\rm circ}\approx{t_{\rm circ}}$ because of $t_X\ll{t}_{\rm circ}$. However, we need to confirm whether there is effectively no continuous super-Eddington accretion onto the SMBH during the circularization through self-consistent radiation hydrodynamic simulations of the circularization process using the energy equation including radiative cooling, advective cooling, and viscous heating. Furthermore, we need to quantify the disk spectrum of the super-Eddington accretion flow based on the slim disk model \citep{1988ApJ...332..646A} for estimating the spectral luminosity of the delayed X-ray flare. We will tackle these problems in the future.

In addition, the viewing angle, which is defined as the angle between the normal to the disk plane and the line of sight, is an important parameter in setting the observed X-ray luminosity. The X-ray luminosity decreases significantly as a function of inclination angle. If the viewing angle goes to zero, the X-ray emission of our fiducial model can reach the luminosity between one-tenth of the Eddington luminosity and the Eddington luminosity. On the other hand, the X-ray luminosity decreases to be less than or approximately equal to one-hundredth of the Eddington luminosity if the viewing angle is close to $90^\circ$. Another important parameter determining the luminosity of the disk black-body radiation is the black hole spin. If the black hole has a maximum positive (prograde) spin, the disk's inner edge is two-thirds smaller than that in the non-rotating case, leading to the larger luminosity in the soft-X-ray waveband (\citealt{1972ApJ...178..347B}, see also figures 2 of both \citealt{2020ApJ...897...80W} and \citealt{2021arXiv210406203M}).

\begin{acknowledgments}
The authors thank the anonymous referee for fruitful comments and suggestions. The authors also thank Nicholas C. Stone for his helpful comments.\,\,The authors acknowledge the Yukawa Institute for Theoretical Physics at Kyoto University, 
where this work was initiated thanks to the YITP-T-19-07 International 
Molecule-type Workshop, "Tidal Disruption Events: General Relativistic Transients".\,\,K.H. has been supported by the Basic Science Research Program through the National Research Foundation of Korea (NRF) funded by the Ministry of Education (2016R1A5A1013277 and 2020R1A2C1007219).
\end{acknowledgments}
%
%
%

%

%

%
%
%
\appendix
%
%


\bibliography{khpj}{}
\bibliographystyle{aasjournal}



\end{document}